\RequirePackage{fix-cm}
\documentclass[smallextended]{svjour3}       
\smartqed  
\usepackage{graphicx}
\usepackage{xcolor}
\usepackage{booktabs}
\usepackage{multirow}
\usepackage{amsmath}
\usepackage{amsfonts}
\usepackage{float}
\usepackage{natbib}
\usepackage{verbatim}
\usepackage{url}

\usepackage{soul}

\usepackage{microtype}

\usepackage{tabularx}

\usepackage{placeins}

\usepackage{mdframed}
\usepackage{xcolor} 

\newcommand{\geribox}[2]{
\par\vspace{10pt}%
\begin{mdframed}[
    userdefinedwidth=0.95\linewidth,
    nobreak=true,
    align=center,
    linewidth=1pt,
    linecolor=gray,
    roundcorner=10pt,
    backgroundcolor=white,
    skipabove=6pt,
    skipbelow=6pt,
    innertopmargin=6pt,
    innerbottommargin=6pt
]
\textit{\textbf{#1}} #2
\end{mdframed}
}

\begin{document}

\title{An Empirical Study on Clustering-based Quantum Approximate Optimization Algorithm for Regression Test Case Selection
}

\titlerunning{An Empirical Study on Clustering-based QAOA for TCS}        

\author{Antonio Trovato \and Martin Beseda \and Dario Di Nucci}
\authorrunning{Trovato \and Beseda \and Di Nucci}  

\institute{
  Antonio Trovato \at
  University of Salerno, Fisciano (SA), Italy \\
  \email{atrovato@unisa.it}
  \and
  Martin Beseda \at
  University of L'Aquila, L'Aquila (AQ), Italy \\
  \email{martin.beseda@univaq.it}
  \and
  Dario Di Nucci \at
  University of Salerno, Fisciano (SA), Italy \\
  \email{ddinucci@unisa.it}
}

\date{Received: date / Accepted: date}

\maketitle

\begin{abstract}
Regression testing assesses whether software components that have already been developed and tested continue to work correctly after modifications. Ideally, regression testing re-executes the entire test suite after each modification, but this is infeasible in real-world scenarios because of the computational burden; therefore, optimized approaches, such as Test Case Selection, have been developed. The state of the art in regression testing optimization includes both classical approaches, such as \textsc{Additional Greedy} and \textsc{DIV-GA}, and quantum approaches, such as \textsc{SelectQA}, \textsc{BootQA}, and \textsc{IGDec-QAOA}. However, while the former incur high computational costs, the latter are constrained by the physical limitations of currently available quantum machines.

This paper extends the evaluation of \textsc{QAOA-TCS}, our quantum approach for the test case selection problem. We thoroughly evaluate its performance, compare it with state-of-the-art classical and quantum approaches, and assess its robustness to quantum noise.

The results show that \textsc{QAOA-TCS} outperforms state-of-the-art multi-objective approaches in most cases and maintains stable performance even under high noise levels. In the single-objective formulation, QAOA-based methods perform comparably. In terms of efficiency, our previous method based on quantum annealing, namely \textsc{SelectQA}, is the most efficient for large problems in the multi-objective formulation, whereas \textsc{QAOA-TCS} offers the best trade-off between effectiveness and efficiency in the single-objective formulation when run on realistic noisy simulators. However, such analysis is constrained by the different computational environments used to run the algorithms.
\end{abstract}


\section{Introduction}
\label{intro}
In modern software development, software systems undergo continuous modifications and enhancements, particularly since the advent of Agile methodologies~\citep {annosi2024agile,meier2023agile}. In this landscape, each modification could interfere with the proper behavior of already developed system components; hence, there is a strong need to continuously monitor software quality. To address this problem, regression testing~\citep{yoo2012regression} verifies, after each developer modification, whether previously implemented and tested components still function as expected, and plays a key role. The ideal regression testing approach, also called \textit{retest-all}, re-executes the entire regression test suite after each software modification; however, in real-world scenarios, it quickly becomes impractical~\citep{rothermel1998empirical,perrouin2012pairwise}. Therefore, many \textit{Regression Test Suite Optimization} (RTSO) methods have been proposed to drastically reduce the costs of regression testing, without losing its effectiveness in finding defects, either by reducing the size of the starting test suite according to specific strategies, or by finding an optimal execution order following some prioritization criteria~\citep{yoo2007pareto,yoo2010using,yoo2011highly,wang2013minimizing,panichella2014improving,arrieta2019pareto,xue2020multi}. Above all, \textit{Test Case Selection} (TCS)~\citep{engstrom2010systematic,yoo2012regression} is one of the most adopted approaches, focusing on identifying a minimal subset of the initial regression suite to minimize overall regression testing cost while preserving effectiveness, as often measured by fault and statement coverage.

Although state-of-the-art classical strategies, such as \textsc{Additional Greedy} \citep{yoo2007pareto} and \textsc{DIV-GA}~\citep{panichella2014improving}, have proven effective for solving TCS, their practical use is limited by their computational burden. Therefore, research has focused on alternatives that leverage the potential of quantum computing, such as superposition and entanglement~\citep{hoare2005grand,knight2018serious}, to overcome the limitations of classical strategies in execution costs without sacrificing solution quality. Currently, such proposed strategies are: \textsc{SelectQA}~\citep{selectqa}, \textsc{BootQA}~\citep{bootqa}, and \textsc{IGDec-QAOA}~\citep{igdec_qaoa}. While \textsc{SelectQA} and \textsc{BootQA} address the TCS problem relying on \textit{Quantum Annealing} (QA)~\citep{kadowaki1998quantum,suzuki2009comparison}, \textsc{IGDec-QAOA} relies on the \textit{Quantum Approximate Optimization Algorithm} (QAOA)~\citep{farhi2014qaoa,qaoa2}. The major limitation of the aforementioned approaches lies in their decomposition strategy; \textsc{BootQA} totally commits to random sampling to produce the sub-suites to optimize, resulting in unstable results; \textsc{IGDec-QAOA} relies on an impact-guided strategy, which does not take into account dependencies among the test cases; finally, \textsc{SelectQA} does not rely on a tailored decomposition strategy at all, thus worsening its efficiency.

This paper extends \citet{qaoa_tcs}, where we proposed \textsc{Quantum Approximate Optimization Algorithm - Test Case Selection (QAOA-TCS)}, a clustering-based QAOA approach for TCS. The novelty of the method was twofold: (i) circumventing the limitations of currently available quantum computers by leveraging a clustering-based decomposition strategy; (ii) applying QAOA to solve the multi-objective TCS problem.

Although our previous study showed that \textsc{QAOA-TCS} outperformed state-of-the-art methods in effectiveness and efficiency, it had several limitations. First, we did not consider the effect of quantum noise in the experiments, since all \textsc{QAOA-TCS} executions were performed on an ideal simulator. Although this choice yielded encouraging preliminary results, it was not sufficient to assess the QAOA-based approach's performance in real environments. Second, although the study compared the proposed approach with well-established state-of-the-art strategies in the literature, it did not consider more recent solutions based on quantum algorithms. 
Therefore, this paper provides a more rigorous and comprehensive empirical analysis of \textsc{QAOA-TCS} compared with state-of-the-art TCS strategies.
Specifically, we have:

\begin{itemize}
    \item Compared \textsc{QAOA-TCS} with state-of-the-art single-objective quantum approaches, which were not included in the original work on \textsc{QAOA-TCS}
    \item Executed both multi- and single-objective versions of our approach over five different noisy backend simulators, namely \textsc{Sampling-noise} simulator, \textsc{Fake Brisbane with sampling noise} simulator, and 1\%, 2\%, and 5\% \textsc{depolarizing-noise models} with sampling noise simulators, to exhaustively evaluate the robustness of \textsc{QAOA-TCS} in noisy quantum environments.
    \item Compared the performance of both the multi-objective and single-objective versions of \textsc{QAOA-TCS} over each noisy environment with the aforementioned classical and quantum state-of-the-art approaches.
    \item Developed an open and reproducible online appendix~\citep{appendix}.
\end{itemize}

The empirical results show that \textsc{QAOA-TCS} outperforms state-of-the-art multi-objective approaches in most cases and maintains stable performance even under high noise levels. In the single-objective formulation, different quantum-based methods perform comparably. Our previous quantum annealing-based method, namely \textsc{SelectQA}, is the most efficient for large problems in the multi-objective formulation. In contrast, \textsc{QAOA-TCS} offers the best trade-off between effectiveness and efficiency in the single-objective formulation when executed on realistic noisy environments. However, such analysis could be hindered by the different computational environments and simulators used to run the algorithms.

\paragraph{Paper Structure.} Section~\ref{sec:2} introduces quantum computing and its noise, details the TCS problem we aim to tackle, and describes the baseline algorithms currently available to solve it, together with their main limitations motivating the need for this study. Section~\ref{sec:3} introduces how \textsc{QAOA-TCS} operates, the mathematical formulations it adopts to represent the TCS problem, and the clustering algorithm used by \textsc{QAOA-TCS} to decompose it. Section 4 describes the design of the empirical evaluation that compares \textsc{QAOA-TCS} with state-of-the-art algorithms, while its threats to validity are provided in Section 6. Finally, Section 7 summarizes the contributions of this work and envisions future work.

\section{Background and Related Work}
\label{sec:2}
This section provides a more detailed introduction to TCS and how state-of-the-art algorithms solve it.

\subsection{Test Case Selection}
\label{sec2.1:tcs}
Test Case Selection (TCS) is a modification-aware optimization strategy that identifies a reduced subset of a regression test suite sufficient to validate whether components of software systems that have already been tested and were indirectly affected by a modification continue to function as expected after the modification~\citep {rothermel1996analyzing}. Several techniques have been proposed to determine which tests are relevant for the selection, including integer programming~\citep{fischer1977test}, symbolic execution~\citep{yau1987method}, data flow analysis~\citep{rothermel1993safe}, dependence graphs~\citep{bates1993incremental}, and control-flow graphs~\citep{rothermel1996analyzing}. Once the candidates are identified, the problem naturally becomes one of optimization: \textit{selecting the smallest possible set of tests while still achieving adequate coverage and fault detection}. The goal is to reduce regression testing costs without compromising its effectiveness.

Although TCS can be treated as a single-objective problem, the natural interplay among its objectives has led researchers to adopt multi-objective formulations. In particular, let $\Gamma = \{\tau_1,\tau_2,\dots,\tau_n\}$ be the starting regression test suite which we aim to optimize; let $F = \{f_1,f_2,\dots,f_m\}$ be the set of objectives of our selection. Multi-objective TCS aims at finding a \textit{Pareto-optimal} set $P$, such that the $\forall p \in P,\; p \subseteq \Gamma$, given the objectives in functions in $F$, where the elements of $P$ are sub-suites (solutions). A solution $p$ belongs to $P$ if it is \textit{non-dominated} by any other solution in the search space; in other words, any other solution $r$ that improves $p$ in some objective function in $F$ worsens it in at least $1$ function. Finally, the Pareto-optimal set consists of all non-dominated solutions, which correspond to points in an $N$-dimensional space ($N=|F|$) that form the \textit{Pareto frontier}.

Leveraging this formulation of TCS, \citet{yoo2007pareto} introduced a multi-objective version of the traditional single-objective greedy strategy, called \textsc{Additional Greedy}. This method produces a solution (i.e., a subsuite) by iteratively selecting test cases that maximize statement and fault coverage over a given time interval. Finally, a Pareto-optimal set is obtained by selecting all non-dominated incremental subsets of the solution set. Nevertheless, this strategy suffers from scalability issues, as its execution time becomes impractical on real-world systems~\citep{greedy_cmplx}.

Therefore, \citet{panichella2014improving} proposed \textsc{DIV-GA}, an approach that improves the established usage of Non-dominated Sorting Genetic Algorithm (\textsc{NSGA-II})~\citep{deb2000nsgaii} by injecting diversity during the evolutionary process. Although \textsc{DIV-GA} was empirically proven to outperform its counterparts \textsc{NSGA-II}, \textsc{Additional Greedy}, and even the \textsc{NSGA-II} variant called \textsc{vNSGA-II}~\citep{yoo2007pareto}, it is limited in terms of execution cost because computing the individuals to inject into populations entails additional computational effort, making this approach impractical in many real-world scenarios.

\subsection{The Impact of the Quantum Era on Test Case Selection}
\label{sec2.2:quantum}

One of the greatest promises of quantum computers is to efficiently solve practically unsolvable optimization problems. The core concept behind this potential is \textit{quantum parallelism}, which exploits the superposition property of quantum bits (qubits), i.e., that a qubit can exist in a state that is a linear combination of 0 and 1, to compute multiple solutions simultaneously.

\textit{Quantum Annealers} are a special kind of quantum machine that execute the \textit{Quantum Annealing} (QA) algorithm, formulated as a Quadratic Unconstrained Binary Optimization (QUBO), as in Eq.~(\ref{eq:qubo}). In QA, a quantum system is initialized so that its ground state can be measured by an easy-to-compute \textit{initial Hamiltonian}, e.g., a mathematical function modeling the energy of a system. QA relies on the \textit{adiabatic theorem}~\citep{kadowaki1998quantum,suzuki2009comparison}, which states that if a quantum system starts in its ground state and its Hamiltonian changes slowly, it remains in the ground state. The system then slowly evolves until its ground state matches the minimal (optimal) value of the \textit{problem Hamiltonian}, i.e., the one encoding the optimization problem we aim to solve.

Building on this paradigm, \citet{bootqa} introduced \textsc{BootQA}, the first approach to apply quantum annealing to test suite optimization. To cope with the limited qubit capacity of current quantum annealers, \textsc{BootQA} relies on \textit{bootstrap sampling}: starting from the original test suite
$TS$, it randomly draws $m$ subsets of size $n$, where $n$ is chosen to fit the hardware constraints. Each subset is independently formulated as a QUBO sub-problem and solved by the annealer; the resulting sub-solutions are then merged into a final solution by retaining all test cases selected across sub-problems.
Despite its efficiency, \textsc{BootQA} has several limitations. First, random sampling does not guarantee that the sampled subsets represent the original suite, leading to unstable coverage and inconsistent solutions. Second, because \textsc{BootQA} samples randomly at each run, result quality depends heavily on the initial random choice. Finally, selecting the optimal hyperparameter pair $(n,m)$ requires repeated trial runs, incurring significant execution time.

In \citet{selectqa}, we proposed \textsc{SelectQA} to address the coverage instability of \textsc{BootQA}. Rather than decomposing the problem manually, \textsc{SelectQA} delegates both decomposition and solving to D-Wave's Hybrid Solver Service\footnote{(Accessed on: 10-06-2026) \url{https://www.dwavesys.com/media/m2xbmlhs/14-1048a-a_d-wave_hybrid_solver_service_plus_advantage_technology_update.pdf}}, which internally combines classical and quantum resources to handle large-scale QUBO instances. We evaluated \textsc{SelectQA} in two variants: a single-objective version directly comparable to \textsc{BootQA} and a multi-objective version compared against classical baselines such as \textsc{Additional Greedy} and \textsc{DIV-GA}. Our experiments showed that \textsc{SelectQA} consistently outperformed all competitors in effectiveness.
However, \textsc{SelectQA} also has limitations. The absence of an explicit decomposition strategy makes it less scalable than \textsc{BootQA} for very large test suites. Furthermore, D-Wave has discontinued the limited free access to its cloud services, making the operational costs of \textsc{BootQA} and \textsc{SelectQA} expensive. Therefore, we investigate an alternative to QA based on quantum gates that addresses the broader physical limitations of quantum computers leveraging problem-decomposition strategies.

\subsection{Gate-based Approaches and Backends}
\label{sec:gate-approaches-backends}
Before introducing the algorithm and the noisy environments, we briefly summarize the concepts underlying
gate-based quantum computing. The fundamental unit of quantum information is a qubit. Unlike a classical bit, which can only assume the values 0 or 1, a qubit can exist in a superposition of basis states,
\begin{align}
|\psi\rangle = \alpha |0\rangle + \beta |1\rangle,
\end{align}
where $\alpha,\beta\in\mathbb{C}$ are probability amplitudes satisfying
\begin{align}
|\alpha|^2+|\beta|^2=1,
\end{align}
as $|\alpha|^2$ and $|\beta|^2$ denote the probability of measuring $|0\rangle$ or $|1\rangle$, respectively.
Another important property is that an isolated quantum system is described by unitary transformations, represented by matrices $U$, thus satisfying
\begin{align}
U^\dagger U = I
\end{align}
and subsequently preserving the norm of the system, i.e., also the probability amplitudes of the normalized quantum state.

When multiple qubits are considered jointly, the state space grows as the tensor product of the individual Hilbert spaces. Consequently, an $n$-qubit state can generally be expressed as \begin{align} 
|\Psi\rangle = \sum_{x=0}^{2^n-1} c_x |x\rangle,
\end{align} 
where $n \in \mathbb{N}$ represents the number of qubits, $|x\rangle$ denotes the computational basis state corresponding to
the binary representation of the $x \in \mathbb{N}$, and
$c_x\in\mathbb{C}$ are probability amplitudes satisfying
\begin{align}
\sum_{x=0}^{2^n-1}|c_x|^2=1.
\end{align}
A multipartite state is called separable if it can be written as
\begin{align} 
|\Psi\rangle = |\psi_1\rangle\otimes|\psi_2\rangle\otimes\cdots\otimes|\psi_n\rangle, 
\end{align} 
where $|\psi_i\rangle$ denotes the state of the $i$-th qubit. Otherwise, the state is entangled. For example, since the Bell state
\begin{align} 
|\Phi^+\rangle = \frac{|00\rangle+|11\rangle}{\sqrt{2}} \end{align} 
cannot be decomposed into a tensor product of two single-qubit states, it represents an entangled state.
Quantum circuits, i.e., sequences of quantum gates that implement unitary or binary operations on qubits, allow quantum algorithms to exploit superposition and entanglement. A quantum circuit implementing a unitary operator $U$ may be expressed as 
\begin{align} 
U = U_pU_{p-1}\cdots U_2U_1, \label{eq:depth}
\end{align} 
where each layer may itself be decomposed as
\begin{align}
\label{eq:layer}
U_d
=
U_d^{(1)}
\otimes
U_d^{(2)}
\otimes
\cdots
\otimes
U_d^{(m_d)},
\end{align}
where $m_d$ denotes the number of gate blocks that can be executed in
parallel within the $d$-th layer. Gates acting on overlapping sets of qubits cannot belong to the same layer and therefore contribute to the circuit depth. The operators $U_d^{(j)}$ act on disjoint subsets of qubits and therefore can be executed in parallel. The quantity $p \in \mathbb{N}$ is the circuit depth and represents the number of sequential gate layers required to implement the algorithm. Circuit depth is an important practical consideration, since gate imperfections and decoherence generally accumulate with the number of sequential operations, making deeper circuits more susceptible to noise.

More recently, these gate-based approaches have gained attention, particularly the \textit{Quantum Approximate Optimization Algorithm} (QAOA)~\citep{farhi2014qaoa}, because gate-based computers are universal, in contrast to quantum annealers, and their universality has been formally proved in~\citep{universalityQC}. The algorithm can be used to solve the problems of \textit{Quadratic Unconstrained Binary Optimization} (QUBO), defined as

\begin{align}
    \min_{x\in \{0, 1\}^k} f(x) = x^TQx,\label{eq:qubo}
\end{align}

where $x = (x_1, \ldots, x_k)^T$ and $x_i\in\{0, 1\}$ and $Q \in \mathbb{R}^{k \times k}$ is a symmetric or upper-triangular matrix, while $k \in \mathbb{N}$. QAOA then relies on the construction of a variational quantum state called ``ansatz'' defined as

\begin{align}
    |\psi(\mathbf{\gamma}, \mathbf{\beta})\rangle = \prod_{j=1}^{p} e^{-i\beta_j B} e^{-i \gamma_j C} |+\rangle^{\otimes k}
\end{align}

where $i$ is an imaginary unit, $p \in \mathbb{N}$ denotes the \textit{depth} of the ansatz according to Eqs.~(\ref{eq:depth}) and (\ref{eq:layer}), $\mathbf{\gamma}, \mathbf{\beta} \in \mathbb{R}^p$,

\begin{align}
|+\rangle^{\otimes k} = \frac{1}{\sqrt{2^k}} \sum_{z_1 = 0}^1\sum_{z_2 = 0}^1\ldots \sum_{z_k = 0}^1 (|z_1\rangle \otimes |z_2\rangle \otimes \cdots \otimes |z_k\rangle), \quad |z_i\rangle \in \{ |0\rangle, |1\rangle\}
\end{align}

and B with C, representing the ``mixer'' Hamiltonian and the ''cost'' Hamiltonian, respectively, are defined as 

\begin{align}
    B &= \sum^k_{i=1} X_i\\
    C &= \sum^k_{i,j = 1} Q_{ij}\frac{(1-Z_i)(1-Z_j)}{4}, 
\end{align}

with 

\begin{align}
    Z_i = I^{\otimes(i-1)} \otimes Z \otimes I^{\otimes(k-i)}.
\end{align}

The cost Hamiltonian $C$ here encodes the QUBO objective function, favoring bitstrings corresponding to better solutions. In contrast, the ``mixer'' Hamiltonian $B$ induces transitions between basis states, enabling exploration of the search space.
Finally, QAOA proceeds by solving the optimization problem

\begin{align}
    \min_{\gamma, \beta} F_p(\mathbf{\gamma}, \mathbf{\beta}) = \langle \psi(\mathbf{\gamma}, \mathbf{\beta}) | C | \psi(\mathbf{\gamma}, \mathbf{\beta}) \rangle
\end{align}

via a classical optimizer.
\smallskip

\textit{Sampling noise} is an inherent quantum-mechanical property, tied to the finite number of measurements we perform. Its effect diminishes as the number of measurements increases due to the \textit{law of large numbers}. Nevertheless, it does not disappear; therefore, we must account for the distortion of the measured probabilities. Assuming a circuit with $n$ qubits, there are $2^n$ possible outcomes following the multinomial distribution given by

\begin{align}
    Pr(\overrightarrow{n}) = \frac{m!}{n_0!n_1!\ldots n_{2^n-1}}\prod_{i=0}^{2^n-1}p_i^{n_i},
\end{align}

where $m=\sum_{i=0}^{2^n-1}n_i$ and $\overrightarrow{n}=(n_0, n_1, \ldots, n_{2^n-1})$ denote the total number of measurements and a specific output vector of measured ``counts''. Defining the measured frequencies as $f_i = \frac{n_i}{m}$, we can describe the standard deviation, i.e., the ``fluctuations'' caused by sampling noise as 

\begin{align}
    \sigma(f_i) = \sqrt{\frac{p_i(1-p_i)}{m}},
\end{align}

thus being of order $\mathcal{O}(1/\sqrt{m})$. The sampling noise was implemented via \texttt{Aer Simulator}\footnote{\url{https://qiskit.github.io/qiskit-aer/stubs/qiskit_aer.AerSimulator.html}} from the Qiskit framework. Because sampling noise is only part of the total noise in real quantum computational infrastructures, we also considered noisy simulators to approximate realistic execution environments. Specifically, the Fake Brisbane backend realistically simulates IBM's Brisbane quantum processor, while depolarizing channels provide simplified, controllable noise models that enable systematic evaluation of QAOA-TCS sensitivity to increasing decoherence levels.
\smallskip

The \textit{Fake Brisbane backend}\footnote{\url{https://quantum.cloud.ibm.com/docs/en/api/qiskit-ibm-runtime/fake-provider-fake-brisbane}} is a complex simulation of quantum decoherence that closely replicates the Brisbane quantum computer environment, providing its coupling map, error channels, and data from the most recent calibration.

\smallskip

The \textit{depolarizing channel} denotes a simplified decoherence model in which, in addition to sampling noise, the qubit state is replaced by the maximally mixed state with probability $p_d$, and it is usually helpful for representing gate errors when the simplification is acceptable. Its effect on the quantum state of one qubit represented via a density matrix $\rho$ can be described as 

\begin{align}
    \varepsilon(\rho) = (1-p_d)\rho + p_d\frac{I}{2}.
\end{align}

This channel was implemented via \texttt{depolarizing\_error()}\footnote{\url{https://qiskit.github.io/qiskit-aer/stubs/qiskit_aer.noise.depolarizing_error.html}} from Qiskit framework.


\vspace{5mm}

In this work, we ran QAOA on an ideal backend and on all three aforementioned backends independently to understand which noise types may make it challenging to apply QAOA-TCS to real-world problems and to identify ways to address them. For the depolarizing channel, we adopted $p_d=\{0.01,0.02,0.05\}$.

\subsection{Differences between IGDec-QAOA and QAOA-TCS}
In the context of gate-based algorithms for TCS, \citet{igdec_qaoa} proposed \textsc{IGDec-QAOA}, the first hybrid classical-quantum approach that combines QAOA with an impact-guided decomposition strategy (\textsc{IGDec}) to tackle single-objective TCO problems. As in all the approaches introduced above, a solution is represented as a bitstring, where the $i-th$ bit, or decision variable, is 1 if the $i-th$ test case is selected, -1 otherwise.

\textsc{IGDec-QAOA} randomly initializes a solution by assigning values to all decision variables and computing the corresponding fitness value $f(x)$. Afterward, it orders test cases by \textit{impact}: for each test case $t_k$, the value of its decision variable $z_k$ is flipped while keeping the others fixed, yielding a new fitness $f'(x)$; the impact of $t_k$ is then defined as $f'(x) - f(x)$,  and test cases are sorted in ascending order of impact. Finally, the sorted test cases are decomposed into fixed-size consecutive groups of $N$ elements, where $N$ depends on the total number of qubits supported by the available quantum hardware, each of which is solved independently by QAOA; the resulting sub-solutions are combined by keeping the optimized values from each group.
This process iterates until a stopping criterion is met, and the best solution encountered across all iterations is returned as the final output.

Despite its contribution, \textsc{IGDec-QAOA} suffers from two main limitations that \textsc{QAOA-TCS} is designed to address. On the one hand, its decomposition is \textit{order-driven but impact-blind to inter-test dependencies}: by selecting consecutive groups of $N$ test cases based on individual impact scores, the strategy does not account for structural relationships among test cases, such as shared statement coverage, which are crucial for TCS. As a consequence, test cases that jointly cover statements may end up in different sub-problems, potentially leading to suboptimal coverage in the merged solution. On the other hand, the quality of the final solution is heavily dependent on the \textit{initial random sampling}: since the impact order is recomputed at each iteration starting from the current solution, a poor random initialization can steer the search towards a suboptimal region of the solution space, from which recovery across iterations is not guaranteed.

\textsc{QAOA-TCS} addresses both limitations through its clustering-based decomposition strategy. By leveraging Ward's hierarchical clustering~\citep{ward1963clustering} on test-case attributes, \textsc{QAOA-TCS} groups structurally similar test cases into the same sub-problem, preserving coverage dependencies throughout the decomposition. Moreover, because clustering is fully deterministic, \textsc{QAOA-TCS} produces stable, reproducible decompositions independent of random initialization, making its results robust across executions. Further details are given in the following section.

\section{QAOA-TCS: a Clustering-based Quantum Approximate Optimization Algorithm for Regression Test Case Selection}
\label{sec:3}

\textsc{QAOA-TCS} applies the Quantum Approximate Optimization Algorithm~\citep{farhi2014qaoa} to solve TCS in both a multi- and single-objective fashion. The core algorithmic pipeline — QUBO formulation, QAOA circuit, and solution merging — was introduced in \citet{qaoa_tcs}. The present paper extends that contribution by: (i) comparing \textsc{QAOA-TCS} with single-objective quantum state-of-the-art approaches by introducing a single-objective version of \textsc{QAOA-TCS}, and (ii) conducting a comprehensive empirical evaluation of both versions under noisy quantum simulators, necessary to assess the \textsc{QAOA-TCS} performance in realistic environments.

\subsection{Overview of the QAOA-TCS Pipeline}

The \textsc{QAOA-TCS} pipeline consists of three main steps, regardless of whether the multi- or single-objective version is used.

\textbf{Step 1 — Clustering-based decomposition.} \textsc{QAOA-TCS} decomposes the initial regression test suite $\Gamma = \{\tau_1, \tau_2, \ldots, \tau_n\}$ into smaller sub-suites. Such decomposition leverages Ward's hierarchical agglomerative clustering method~\citep{ward1963clustering}. Using a feature vector, we represent test-case attributes relevant to the optimization objectives, standardized via $z$-score normalization (e.g., execution cost, failure rate, and statement coverage). The resulting feature vector is a composite, heterogeneous representation of test cases that can be adapted to any combination of TCS objectives available in a given dataset. Ward's method minimizes the total within-cluster variance at each merge step, grouping test cases similar across all considered attributes.

\textbf{Step 2 — QAOA optimization of sub-problems.} Each cluster of test cases is independently formulated as a QUBO problem and directly solved by QAOA. Each sub-problem involves at most $N$ binary decision variables $x_i \in \{0, 1\}$, where $x_i = 1$ indicates that test case $\tau_i$ is selected, 0 otherwise. The solution to each sub-problem is a bitstring of length equal to the cluster size, encoding the selection decisions for the test cases in that cluster.

\textbf{Step 3 — Solution merging.} The sub-solutions produced in Step 2 are merged into a single final solution by taking their union: a test case $\tau_i$ is included in the final suite if and only if it was selected ($x_i = 1$) in at least one sub-solution. In the multi-objective version, a Pareto frontier is then constructed from the final solution using the same strategy as \textsc{Additional Greedy}~\citep{yoo2007pareto}.

\subsection{Why Does Clustering Improve QAOA-TCS Beyond Hardware Constraints?}

Ward's hierarchical clustering serves a dual purpose as a decomposition strategy in \textsc{QAOA-TCS}.

On the one hand, current quantum hardware makes the problem solvable only after partitioning the test suites into bounded-size subproblems. This constraint is shared by all existing quantum TCS approaches.

On the other hand, grouping test cases with similar optimization-relevant attributes ensures that the test cases most likely to interact are co-located within the same sub-problem. In other words, when solving each sub-problem, the QAOA optimizer can reason jointly over test cases that share statement coverage, exhibit similar failure rates, or incur comparable execution costs. As a consequence, the selection decisions made within each cluster are locally coherent with respect to the TCS objectives, and the merged final solution inherits this coherence globally.

This strategy contrasts with the decomposition strategies of \textsc{BootQA} and \textsc{IGDec-QAOA}, which assign test cases to sub-problems without analyzing attribute similarity. In \textsc{BootQA}, the assignment is purely random, and test cases sharing statement coverage may end up in different sub-problems, causing redundant selections in the merged solution. In \textsc{IGDec-QAOA}, the assignment is driven by individual impact scores, which capture each test case's marginal contribution in isolation but ignore the joint interactions among test cases. By contrast, the clustering-based decomposition of \textsc{QAOA-TCS} aligns the sub-problem boundaries with the natural structure of the TCS objective, enabling QAOA to exploit its quadratic optimization capacity on the most relevant groups of test cases.

The following subsections present the mathematical formulations for both versions, inspired by~\cite{serrano2024incident}, where TCS is formulated and solved as the \textit{Minimum Vertex Cover}. Table~\ref{tab:1} summarizes all definitions needed to understand such formulations.

\begin{table}[ht]
\caption{Definitions of Variables}
\label{tab:1}       
\begin{tabularx}{\linewidth}{lX}
\hline
\textbf{\#} & \textbf{Definition} \\
\hline
$i$ & index identifying a test case \\
$k$ & index identifying a statement \\
$\Gamma$ & the initial regression test suite \\
$T_k$ & list of test cases executing the $k$-th statement \\
$x_i$ & binary variable specifying whether the $i$-th test case has been selected \\
$cost(\tau_i)$ & (normalized) execution cost of the $i$-th test \\
$e_i$ & binary variable specifying whether the $i$-th test case detected faults in the past \\
$f_i$ & the failure rate, i.e., the percentage of times a test detected a fault in the past \\
$d_i$ & the $i$-th test case diversity \\
$p_i$ & (normalized) passenger count of the $i$-th test case \\
$t_i$ & (normalized) travel distance of the $i$-th test case \\
$\alpha$ & weight factor modulating the contrast ratio between the objectives of the problem \\
$P$ & penalty coefficient modeling the importance of constraints of the problem when searching for the solution \\
\hline
\end{tabularx}
\end{table}

\subsection{Multi-Objective QAOA-TCS}
\label{sec:3.1.1}
The multi-objective formulation of \textsc{QAOA-TCS} follows the mathematical model introduced in the multi-objective version of \textsc{SelectQA}~\citep{selectqa}, which selects the smallest subset of test cases that preserves statement and fault coverage. Specifically, we considered \emph{execution cost}, \emph{statement coverage}, and \emph{fault coverage}, in line with the other baselines we considered: \textsc{SelectQA}~\citep{selectqa}, \textsc{Additional Greedy}~\citep{yoo2007pareto}, and \textsc{DIV-GA}~\citep{panichella2014improving}.  

As in \textsc{SelectQA}, each test case is represented by a binary decision variable $x_i \in \{0,1\}$, which indicates whether the test case $\tau_i$ is included in the selected sub-suite. The optimization problem is formulated as a Quadratic Unconstrained Binary Optimization (QUBO) problem, where the linear terms represent the contributions of each test case to execution cost and fault coverage, and the off-diagonal quadratic terms capture interactions among test cases through overlapping statement coverage. The final QUBO integrates: (i) minimization of the total execution cost of selected tests, (ii) maximization of fault detection ability, and (iii) maximization of statement coverage.
Formally, it is the same as that used by \textsc{SelectQA}, which is defined as

\begin{equation}
H = \alpha \sum_{i=1}^{|\Gamma|} x_i \cdot cost(\tau_i) 
  - (1-\alpha)\sum_{i=1}^{|\Gamma|} (e_i \cdot x_i) 
  + \sum_{k}\sum_{i,j \in T_k} (-P x_i + 2 P x_i x_j)
\end{equation}

where $\alpha$ is a weight factor regulating the trade-off between objectives, and $P$ is a penalty coefficient enforcing statement coverage constraints.

\subsection{Single-Objective QAOA-TCS}
\label{sec:3.1.2}

We leveraged the TCS QUBO of \textsc{SelectQA}~\citep{selectqa}, specifically its single-objective version in this case, also to compare \textsc{QAOA-TCS} with single-objective approaches such as \textsc{SelectQA}~\citep{selectqa}, \textsc{BootQA}~\citep{bootqa}, and \textsc{IGDec-QAOA}~\citep{igdec_qaoa}. The general idea is to aggregate multiple criteria into a weighted linear function to minimize. 
The objective function is expressed as 

\begin{equation}
H = \alpha \sum_{i=1}^{|\Gamma|} x_i \cdot cost(\tau_i) 
  - (1-\alpha) \sum_{i=1}^{|\Gamma|} (f_i \cdot x_i),
\end{equation}

where $cost(\tau_i)$ indicates the normalized \emph{execution cost} of the test case $\tau_i$, and $f_i$ is its \emph{failure rate}, i.e., the rate $\tau_i$ has revealed faults in the past.  
Since \textsc{IGDec-QAOA} defines two additional formulations to optimize a test suite for an elevator system, we introduce the corresponding formulations for \textsc{QAOA-TCS}.
The first one differs from the previous one by substituting the failure rate objective with \emph{test case diversity}, denoted as $d_i$, thus describing the objective function as:

\begin{equation}
H = \alpha \sum_{i=1}^{|\Gamma|} x_i \cdot cost(\tau_i) 
  - (1-\alpha) \sum_{i=1}^{|\Gamma|} (d_i \cdot x_i).
\end{equation}

In the third formulation, the weighted sum combines the \emph{execution cost} objective with \emph{passenger count} $p_i$ and \emph{travel distance} $t_i$, as

\begin{equation}
H = \alpha \sum_{i=1}^{|\Gamma|} x_i \cdot cost(\tau_i) 
  - \beta \sum_{i=1}^{|\Gamma|} (p_i \cdot x_i)
  - \gamma \sum_{i=1}^{|\Gamma|} (t_i \cdot x_i),
\end{equation}

where $\alpha$, $\beta$, and $\gamma$ are weight coefficients that balance the three objectives.

\section{Experimental Design}
\label{sec:design}

This section presents the experimental design of our empirical evaluation of QAOA-TCS against the considered baselines, including its goal and research questions. The first part focuses on the multi-objective formulation of \textsc{QAOA-TCS}, whereas the second focuses on the single-objective one. For each formulation, it introduces the test suite datasets used as the basis for the selection problem, the algorithm configurations used in the experiments, and the metrics and statistical tests used to quantitatively measure and reliably discuss the results.

\subsection{Research Goals and Research Questions}
\label{sec:4.1}
The main \textit{objective} of this evaluation is to assess the performance of \textsc{QAOA-TCS} on ideal and noisy environment simulators, compared with both classical and quantum state-of-the-art methods, in terms of effectiveness and efficiency. Our results are expected to be of interest to both practitioners and researchers: the former seek a \textit{cost-effective} strategy to reduce the time to market of their products by speeding up the testing activities; the latter, especially in the field of quantum computing for software engineering, are concerned with continuously enhancing the state-of-the-art. To this aim, we define the following research questions:  

\begin{description}
    \item[\textbf{RQ$_1$.}] To what extent does the circuit depth of QAOA impact the effectiveness of \textsc{QAOA-TCS} in its multi- and single- objective formulations?
    \item[\textbf{RQ$_2$.}] What is the effectiveness of \textsc{QAOA-TCS} compared with state-of-the-art approaches in its multi- and single- objective formulations when executed on an ideal simulator?
    \item[\textbf{RQ$_3$.}] How robust is QAOA-TCS to noisy quantum environments in its multi- and single- objective formulations?
    \item[\textbf{RQ$_4$.}] How efficient is \textsc{QAOA-TCS} compared with state-of-the-art approaches in its multi- and single- objective formulations?
\end{description}

Since we aim to evaluate both multi- and single-objective TCS formulations, we use distinct metrics and datasets for each.

\subsection{Evaluating the Multi-Objective QAOA-TCS}
Next, we present the experimental design to compare multi-objective QAOA-TCS with state-of-the-art multi-objective test case selection approaches.

\subsubsection{Study Design}
\label{sec:4.2.1}

\paragraph{Study Context.}
To select the systems under test (SUTs) for our evaluation of multi-objective \textsc{QAOA-TCS}, we relied on the \textit{Software-artifact Infrastructure Repository} (SIR)\footnote{(Accessed on: 12-17-2025) https://sir.csc.ncsu.edu/portal/index.php}. Specifically, we have chosen four systems, namely \textit{Flex}, \textit{Grep}, \textit{Gzip}, and \textit{Sed}. We chose SIR programs because they have been widely used in prior studies~\citep{selectqa,yoo2007pareto,panichella2014improving} on this topic, particularly for evaluating our baselines.
Table~\ref{table:sir_programs} summarizes the key characteristics of such programs.

\begin{table}[h]
\footnotesize
\centering
\caption{Characteristics of the chosen SIR programs~\citep{qaoa_tcs}}
\label{table:sir_programs}
\begin{tabular}{@{}lrrr@{}}
\hline\noalign{\smallskip}
\textbf{Program} & \textbf{LOC} & \textbf{\# TCs} & \textbf{Description} \\
\noalign{\smallskip}\hline\noalign{\smallskip}
Flex    & 10,459   & 567  & Fast lexical analyzer  \\
Grep    & 10,068   & 808  & Regular expression utility  \\
Gzip    & 5,680    & 215  & Data compression program  \\
Sed     & 14,427   & 360  & Non-interactive text editor  \\
\noalign{\smallskip}\hline
\end{tabular}
\end{table}

To collect the data required for the three objectives, we adopted the same methodology we previously employed for \textsc{SelectQA} and \textsc{DIV-GA}~\citep{selectqa,panichella2014improving}:

\begin{itemize}
\item \textit{Statement Coverage.} Through \textit{gcov}, we could get the statements covered by each executed test case.

\item \textit{Execution Cost.} Following \citet{panichella2014improving}, we used \textit{gcov}, considering the execution cost to be the execution frequency of basic blocks, i.e., elementary instructions of the code, which provides a finer granularity than source lines.

\item \textit{Past Faults Coverage.} The SIR repository provides multiple faulty versions for each program, together with a \textit{fault matrix} indicating whether each test case revealed an injected fault, which we have converted into a bitstring, having the $i-th$ bit corresponding to the $i-th$ test case.
\end{itemize}

\paragraph{Experiment Configurations.}
\label{4.2.2}
We compared the multi-objective \textsc{QAOA-TCS} with \textsc{Additional Greedy}~\citep{yoo2007pareto}, \textsc{DIV-GA}~\citep{panichella2014improving}, and the multi-objective \textsc{SelectQA}~\citep{selectqa}.
To ensure robust empirical results, we ran all stochastic algorithms ten times, consistent with \citet{bootqa,selectqa}, whereas \textsc{Additional Greedy}, being fully deterministic, was run only once.
To address RQ3, we ran \textsc{QAOA-TCS} ten times for each simulator.

Before comparing \textsc{QAOA-TCS} with multi-objective state-of-the-art approaches, we tuned the circuit depth hyperparameter of QAOA by executing \textsc{QAOA-TCS} on the ideal simulator ten times for each SIR program, varying the circuit depth in the $\{1,2,4,8,16\}$ range, following \citet{igdec_qaoa}.
We ran \textsc{Additional Greedy} and \textsc{SelectQA} under the same conditions as in \citet{selectqa}; we implemented \textsc{DIV-GA} using the \textit{MATLAB R2024a Global Optimization Toolbox}, following the setup in \citet{panichella2014improving}.
\textsc{QAOA-TCS} was executed on six different backends provided by \textit{Qiskit}\footnote{(Accessed On: 01-11-2026): https://www.ibm.com/quantum/qiskit}: (i) the ideal noiseless \textit{Statevector Simulator}, which has been employed also by \textsc{IGDec-QAOA}~\citep{igdec_qaoa}, (ii) the noisy simulator modeling sampling noise only, (iii) the ``realistic'' \textit{Fake Brisbane} backend based on IBM's Brisbane computer, chosen to be coherent with \citet{igdec_qaoa}, and  (iv–vi) three noisy simulators with depolarization error rates of 1\%, 2\%, and 5\%, each combined with sampling noise.
We used the Qiskit Aer framework to integrate QAOA with the aforementioned simulators.
All implementations and configurations are available in our replication package~\citep{appendix}.

\paragraph{Evaluation Metrics (Effectiveness).}
\label{sec:4.2.3}  
As in \citet{selectqa,panichella2014improving}, the quality of a multi-objective optimization approach was evaluated using a \textit{reference Pareto frontier}. Since computing the actual frontier is impractical for large-scale problems, we constructed an \textit{a posteriori} reference frontier by aggregating all non-dominated solutions produced across all algorithms and runs. Formally, given $P=\{P_1,\ldots,P_l\}$ as the set of $l$ different Pareto frontiers generated by each algorithm during each run, the reference frontier $P_{ref}$ is defined as:

\begin{equation}
P_{ref} \subseteq \bigcup_{i=1}^{l} P_i
\end{equation}

where $\forall p \in P_{ref}, \not\exists q \in P_{ref} : q > p$.  

To practically measure the effectiveness of the approaches, we computed the \textit{number of non-dominated solutions} contributed by the Pareto frontier of each algorithm to the reference Pareto frontier, in line with \citet{panichella2014improving} and \citet{selectqa}. In other words, for each frontier built by an algorithm, we computed the number of solutions on that frontier that are non-dominated by any other solution produced by all the other approaches.

To further analyze the effectiveness of the Pareto frontier, we have included the computation of two further metrics used in the context of multi-objective algorithms~\citep{nery2015hypervolume,auger2009hypervolume,moffaert2016hypervolume}: Hypervolume (HV), measuring the volume of the objective space dominated by a Pareto frontier $P$ and bounded by a reference point $r$, and Inverted Generational Distance (IGD), which measures how far a Pareto frontier approximation $P$ is from a Reference one $R$. Specifically, by setting the reference point $r$ as $(1.1,1.1,1.1)$, chosen since it is slightly worse than the nadir point~\citep{ishibuchi2018hypervolume}, and using the $P_{ref}$ frontier as the reference one, we have computed the two metrics for each Pareto frontier computed by each algorithm execution, resulting in ten HV and ten IGD values for each algorithm.

\paragraph{Evaluation Metrics (Efficiency).} To compare the efficiency of each adopted algorithm, we have computed their mean run time over ten executions for each considered program. All algorithms, except for \textsc{SelectQA} which, following \citet{selectqa}, was executed on the D-Wave Leap Hybrid Solver \texttt{hybrid\_binary\_quadratic\_model\_version2}\footnote{(Accessed on: 10-06-2026) \url{https://www.dwavesys.com/media/m2xbmlhs/14-1048a-a_d-wave_hybrid_solver_service_plus_advantage_technology_update.pdf}}, have been deployed on the same Windows 11 machine providing 32GB of RAM, Intel(R) Core(TM) Ultra 9 285K CPU of 3.70GHz, and a NVIDIA GeForce RTX 5090 GPU with 31.50GB. Since the decomposition time of \textsc{QAOA-TCS} was always under $0.01s$, we excluded it from the evaluation.

\paragraph{Statistical Analysis.}
To assess the reliability of the observed results, we statistically analyzed each metric: the number of non-dominated solutions, HV, and IGD. Since each approach, except \textsc{Additional Greedy}, was executed ten times per SIR program, we obtained ten values per metric per approach. We first applied the \textit{Shapiro-Wilk normality test}~\citep{wilkshapiro} to each group of ten values to determine the appropriate statistical test. Based on its outcome, for each program and metric, we followed one of two paths:

\begin{itemize}
    \item If at least one group was non-normally distributed, we applied the \textit{Kruskal\-–Wallis H test}~\citep{h_test} with null hypothesis $H_0$: \textit{``There is no statistically significant difference between the groups.''}  When significance was detected ($p$-value $< 0.05$), we performed \textit{post-hoc Dunn's tests}~\citep{dunn_test} with \textit{Bonferroni} correction~\citep{weele2019bonferroni}, with null hypothesis $H_0$: \textit{``There is no statistically significant difference between the pair.''} Effect size was quantified via the \textit{Vargha–Delaney} $\hat{A}_{12}$ statistic~\citep{a12}.
    \item If all groups were normally distributed, we applied a one-way \textit{ANOVA}~\citep{anova} test with the same null hypothesis as the H test. When significance was detected, we performed \textit{post-hoc Tukey HSD}~\citep{tuckey} pairwise comparisons with \textit{Bonferroni} correction using the same null hypothesis as Dunn's test, and quantified effect size via \textit{Cohen's d}~\citep{conover1998cohen}.
\end{itemize}

Following \citet{selectqa}, $\hat{A}_{12}$ values are labeled \textit{small} if in (0.34,0.44] or [0.56, 0.64), \textit{medium} if in (0.29,0.34] or [0.64,0.71), and \textit{large} if in [0,0.29] or [0.71,1]. The labeling for Cohen's $d$ follows~\citep{panichella2014improving} thresholds: \textit{small} $|d| \in (0.2, 0.5]$, \textit{medium} $|d| \in (0.5, 0.8]$, and \textit{large} $|d| > 0.8$.\\

All results and implementations are provided in our replication package~\citep{appendix}.


\begin{table}[ht]
\footnotesize
\centering
\caption{Characteristics of the evaluated datasets}\label{table:datasets}
\begin{tabular}{@{}lrr@{}}
\hline\noalign{\smallskip}
\textbf{Dataset} & \textbf{\# TCs} & \textbf{Description} \\
\noalign{\smallskip}\hline\noalign{\smallskip}
GSDTSR    & 5,553   & Google shared dataset of test suite results \\
Paint Control    & 89   & Painting robot test suite \\
IOF/ROL    & 1,941   & Robot test suite \\
Elevator    & 1,925   & Test elevator scheduler \\
\noalign{\smallskip}\hline
\end{tabular}
\end{table}

\subsection{Evaluating the Single-Objective QAOA-TCS}
\label{sec:4.3}
Next, we present the experimental design to compare single-objective QAOA-TCS with state-of-the-art single-objective test case selection approaches.

\subsubsection{Study Design}
\label{sec:4.3.1}

\paragraph{Study Context.}
The \textit{context} of this part of the study, depicted by Table~\ref{table:datasets}, extends the experimental setup used in~\citep{selectqa} and~\citep{bootqa}. In addition to the datasets we previously considered in~\citep{selectqa}, namely \textit{Paint Control} from ABB Robotics Norway and \textit{GSDTSR} from Google, whose test cases are characterized by \emph{execution cost} and \emph{failure rate}, we included two further datasets used in the evaluation of \textsc{IGDec-QAOA}~\citep{igdec_qaoa}.  

The \textit{IOF/ROL} dataset provided by ABB Robotics Norway contains 1,941 tests characterized by \textit{execution time} and \textit{failure rate}.
The \textit{ELEVATOR} dataset, from Orona5, comprises a test suite of 1,925 elements that test an elevator-scheduling dispatcher. Since this dataset comes with different attributes for its test cases, it also enables the formalization of two different optimization scenarios: (i) \textit{ELEVATOR$_{o2}$}, in which the test cases are characterized by execution cost and input diversity; (ii) \textit{ELEVATOR$_{o3}$}, characterizing test cases with execution cost, passenger counts, and travel distance.
Following \citet{bootqa,selectqa,igdec_qaoa}, we filtered out all test cases with a zero failure rate from Paint Control, GSDTSR, and IOF/ROL.

\paragraph{Experiment Configuration.}
To account for the stochastic nature of quantum algorithms and obtain statistically reliable results, we executed each single-objective approach ten times, in line with \citet{bootqa,selectqa}.  
\textsc{BootQA} was cloned from its GitHub page\footnote{(Accessed on: 12-17-2025) https://github.com/AsmarMuqeet/BootQA} and executed in its optimal configurations of $(n,m)$ parameters, as empirically determined in \citet{bootqa}. Similarly, we ran \textsc{SelectQA} using the same procedure described in \citet{selectqa} to ensure full comparability with the reported results.  
New time limits for experimenting with D-Wave quantum annealers prevented us from running \textsc{BootQA} and \textsc{SelectQA} on the IOF/ROL and ELEVATOR datasets; therefore, we compared these approaches only against the QAOA approaches on the GSDTSR and Paint Control datasets.

We executed \textsc{IGDec-QAOA} as described in \citet{igdec_qaoa}, using both the ideal simulator and the \textit{Fake Brisbane} backends, with ten repetitions for each configuration and a QAOA depth of one. The problem decomposition size was fixed at seven, corresponding to the optimal setting identified in \citet{igdec_qaoa}.
The single-objective \textsc{QAOA-TCS} was also executed ten times using the same ideal and noisy simulators as in the multi-objective case. We also set the number of sub-problems obtained via clustering to 7, consistent with the sub-problem size of \textsc{IGDec-QAOA}.
To compute the optimal QAOA circuit depth, we ran \textsc{QAOA-TCS} on the ideal simulator for each dataset ten times across depths $\{1,2,4,8,16\}$. Once the optimal value was determined, it was used to configure \textsc{QAOA-TCS} for comparisons with the other approaches.
Because both \textsc{IGDec-QAOA} and \textsc{QAOA-TCS} let developers adjust the relative importance of individual selection criteria through objective-specific weights, we used the Optuna framework~\citep{optuna_2019} to examine how these weights affect the algorithms' performance. With Optuna, we systematically explored the weights' configuration space, treating them as hyperparameters of the test case selection process.
All implementations and configurations are included in our replication package~\citep{appendix}.

\paragraph{Evaluation Metrics (Effectiveness).}
\label{sec:eval_metrics_2}
Since none of the algorithms shared the same objective function, we could not directly compare the solutions produced by the approaches.
For the \textit{Paint Control}, \textit{GSDTSR}, and \emph{IOF/ROL} datasets, effectiveness was measured by the execution time and failure rate of the resulting suites, consistent with \citet{bootqa,selectqa}.  
For the other two scenarios, we used execution time and input diversity for \emph{ELEVATOR$_{o2}$}, and execution time, passenger counts, and travel distance for \emph{ELEVATOR$_{o3}$}.

\paragraph{Evaluation Metrics (Efficiency).} To compare the approaches' efficiency, we collected their mean run time for each experiment repetition over the considered datasets. We ran \textsc{SelectQA} and \textsc{QAOA-TCS} on the same hardwares used for them in the multi-objective setting, while \textsc{IGDec-QAOA} was executed on the same Windows 11 machine used for \textsc{QAOA-TCS}. Furthermore,  \textsc{BootQA} was executed on the D-Wave \textit{Advanced Systems QPU}\footnote{(Accessed on: 10-06-2026): \url{\url{https://www.dwavesys.com/media/s3qbjp3s/14-1049a-a_the_d-wave_advantage_system_an_overview.pdf}}}, following \citet{bootqa}. Since the decomposition times of \textsc{QAOA-TCS}, \textsc{IGDec-QAOA}, and \textsc{BootQA} are all under the $0.01s$, we excluded them from the evaluations.

\paragraph{Statistical Analysis.}
Because we quantify the effectiveness of the approaches using different metrics across datasets (Section~\ref{sec:eval_metrics_2}), we performed the statistical analysis separately for each dataset--metric pair.
Specifically, for each dataset and each metric, we first assessed whether the available data could be treated as following a normal distribution; based on this first evaluation, we chose the appropriate parametric or non-parametric tests to assess whether our findings were statistically grounded.

To decide whether to use parametric or non-parametric tests, we applied the \textit{Shapiro-Wilk normality test} to each data group.
Finally, for each dataset, we applied the following strategy to the sets of ten results for each metric and approach.

\begin{itemize}
    \item To analyze the effect of QAOA circuit depth on \textsc{QAOA-TCS}, we used \textit{ANOVA} with \textit{Tukey}’s post-hoc test and Cohen's $d$ effect size with the null hypothesis $H_0$: \textit{``There is no statistically significant difference between the groups of data.''}
    \item We compared \textsc{QAOA-TCS} and \textsc{IGDec-QAOA} on \textit{IOF/ROL}, \textit{ELEVATOR$_{o2}$}, and \textit{ELEVATOR$_{o3}$} using the \textit{Mann–Whitney U test} and $\hat{A}_{12}$ with the null hypothesis $H_0$: \textit{``There is no statistically significant difference between the groups of data.''}
    \item For the \textit{GSDTSR} dataset, we used the \textit{Kruskal–Wallis H test} with Bonferroni-corrected Dunn’s test and $\hat{A}_{12}$ with the null hypothesis $H_0$: \textit{``There is no statistically significant difference between the groups of data.''}
    \item For the \textit{Paint Control} dataset, we used \textit{ANOVA} with \textit{Tukey's test} and Cohen's $d$ and \textit{Kruskal–Wallis H test} with Bonferroni-corrected \textit{Dunn's test} and $\hat{A}_{12}$ for comparing the failure rates and the execution costs, respectively, with the null hypothesis $H_0$: \textit{``There is no statistically significant difference between the groups of data.''}
\end{itemize}

We interpreted the effect size values as we did in the analysis of the multi-objective QAOA-TCS version.
The replication package~\citep{appendix} includes the implementation of the statistical tests, along with tables showing the results of all statistical analyses.

\section{Empirical Evaluations}
\label{sec:results}
This section presents the experimental results, including comparisons between algorithms and key findings, and offers insights for practitioners and researchers. For the sake of space, all the tables depicting the results of the statistical tests have been left in the online appendix~\citep{appendix}.

\subsection{Experimental Results of Multi-Objective Algorithms}
This section presents and discusses the results of the comparisons of the multi-objective algorithms for TCS.

\subsubsection{RQ$_{1.1}$: To what extent does the circuit depth of QAOA impact the effectiveness of QAOA-TCS in its multi-objective formulation?}
The results of the Kruskal-Wallis H test show no significant differences between the values of non-dominated solutions among the results obtained by \textsc{QAOA-TCS} executed over the different configurations, with all $p-$values exceeding the $0.05$ threshold. This result indicates that, for each SIR program, increasing the depth of the QAOA circuit does not influence the quality of \textsc{QAOA-TCS} results; therefore, we tuned \textsc{QAOA-TCS} in the following analyses with a depth level of $p=1$, as it minimizes the computational cost of QAOA without compromising the effectiveness of our approach.

\geribox{Takeaway \#1.}{Circuit depth does not impact the performance of QAOA-TCS in a statistically significant manner. Configuring the circuit depth to $p=1$ is the most appropriate choice, as it minimizes the computational cost of QAOA without compromising the effectiveness of our approach.}

\subsubsection{RQ$_{2.1}$. What is the effectiveness of multi-objective \textsc{QAOA-TCS} compared with state-of-the-art approaches when executed on an ideal simulator?}

This section reports results obtained by QAOA-TCS on the ideal simulator and compares them with the baseline approaches.

\label{sec:5.2.1}
\begin{table*}[t]
\centering
\scriptsize
\setlength{\tabcolsep}{3pt}
\caption{Mean and standard deviation of non-dominated solutions, HV, and IGD for QAOA-TCS compared to state-of-the-art methods on the Ideal Simulator.}
\label{tab:ideal_means}
\resizebox{\textwidth}{!}{%
\begin{tabular}{llccc}
\toprule
\textbf{Program} & \textbf{Algorithm} & \textbf{Non-dom. sol.} $\uparrow$ & \textbf{HV} $\uparrow$ & \textbf{IGD} $\downarrow$ \\
\midrule
Flex & Additional Greedy & 199.0 & 0.3995 & 0.3872 \\
 & SelectQA & 187.0 & 0.4622 & 0.3313 \\
 & DIV-GA & 139.7 $\pm$ 0.67 & 0.6668 $\pm$ 0.0103 & 0.2148 $\pm$ 0.0077 \\
 & QAOA-TCS & \textbf{229.1 $\pm$ 3.38} & \textbf{0.9599 $\pm$ 0.0075} & \textbf{0.0857 $\pm$ 0.0010} \\
\midrule
Grep & Additional Greedy & 167.0 & 0.3577 & 0.4082 \\
 & SelectQA & 9.5 $\pm$ 0.53 & 0.3610 & 0.2553 \\
 & DIV-GA & 70.0 & 0.5790 $\pm$ 0.0227 & 0.3309 $\pm$ 0.0131 \\
 & QAOA-TCS & \textbf{326.2 $\pm$ 3.08} & \textbf{0.7907 $\pm$ 0.0055} & \textbf{0.0758 $\pm$ 0.0013} \\
\midrule
Gzip & Additional Greedy & 72.0 & 0.3348 & 0.3773 \\
 & SelectQA & 38.3 $\pm$ 0.82 & 0.4380 & 0.2999 \\
 & DIV-GA & \textbf{104.9 $\pm$ 0.32} & \textbf{0.6103 $\pm$ 0.0257} & 0.2804 $\pm$ 0.0171 \\
 & QAOA-TCS & 89.5 $\pm$ 0.53 & 0.4828 $\pm$ 0.0031 & \textbf{0.2567 $\pm$ 0.0025} \\
\midrule
Sed & Additional Greedy & 83.0 & 0.4744 & 0.2359 \\
 & SelectQA & 83.0 & 0.3868 & 0.3011 \\
 & DIV-GA & 99.2 $\pm$ 9.54 & 0.5831 $\pm$ 0.0140 & 0.2495 $\pm$ 0.0103 \\
 & QAOA-TCS & \textbf{120.6 $\pm$ 3.10} & \textbf{0.7221 $\pm$ 0.0056} & \textbf{0.0582 $\pm$ 0.0016} \\
\bottomrule
\end{tabular}%
}
\end{table*}

We compared \textsc{QAOA-TCS} executed on the ideal simulator against the state-of-the-art multi-objective approaches. As shown in Table~\ref{tab:ideal_means}, \textsc{QAOA-TCS} achieves the best overall performance on \textit{Flex}, \textit{Grep}, and \textit{Sed}, reporting the largest number of non-dominated solutions, the highest HV, and the lowest IGD. The statistical analysis mostly confirms this trend; in terms of non-dominated solutions, \textsc{QAOA-TCS} significantly outperforms \textsc{DIV-GA} and \textsc{SelectQA} on \textit{Flex} and \textit{Grep}, with large effect sizes, while on \textit{Sed}, \textsc{QAOA-TCS} is significantly better than \textsc{SelectQA} and \textsc{Additional Greedy}, always with large effect sizes. For HV and IGD, improvements are statistically significant across all baselines on \textit {Sed}, while on \textit{Flex} and \textit{Grep} only the comparisons with \textsc{Additional Greedy} are significant. In \textit{Gzip}, \textsc{DIV-GA} obtains a larger number of non-dominated solutions and a higher HV than \textsc{QAOA-TCS}; however, the difference in the number of non-dominated solutions is not statistically significant, while \textsc{QAOA-TCS} still obtains statistically significantly better results in terms of IGD values than \textsc{DIV-GA}.

\geribox{Takeaway \#2.}{\textsc{QAOA-TCS} performs statistically better than the multi-objective baseline approaches in most cases; only in one case did \textsc{DIV-GA} obtain better results, but these findings are not statistically significant in terms of non-dominated solutions.}

\subsubsection{RQ$_{3.1}$. How robust is QAOA-TCS to noisy quantum environments in its multi-objective formulation?}

For each noisy configuration, Tables \ref{tab:samplenoise_means} to \ref{tab:noise5_means} report the absolute values of the three adopted metrics and include, in italics, the corresponding \textsc{QAOA-TCS} result obtained on the ideal simulator, allowing us to compare with state-of-the-art approaches and assess the degradation or preservation of \textsc{QAOA-TCS} compared to its ideal execution.

\begin{table*}[ht]
\centering
\scriptsize
\setlength{\tabcolsep}{3pt}
\caption{Mean and standard deviation of non-dominated solutions, HV, and IGD for QAOA-TCS compared to state-of-the-art methods on the simulator with sampling noise. The italic row reports the Ideal QAOA-TCS reference.}
\label{tab:samplenoise_means}
\begin{tabular}{llccc}
\toprule
\textbf{Program} & \textbf{Algorithm} & \textbf{Non-dom. sol.} $\uparrow$ & \textbf{HV} $\uparrow$ & \textbf{IGD} $\downarrow$ \\
\midrule
Flex & Additional Greedy & 1.0 & 0.3995 & 0.3024 \\
 & SelectQA & 187.0 & 0.4622 & 0.3686 \\
 & DIV-GA & 139.4 $\pm$ 0.70 & 0.6668 $\pm$ 0.0103 & 0.2146 $\pm$ 0.0084 \\
 & QAOA-TCS & \textbf{241.8 $\pm$ 63.90} & 0.6966 $\pm$ 0.1926 & 0.1372 $\pm$ 0.0116 \\
 & \textit{QAOA-TCS (Ideal)} & \textit{229.1 $\pm$ 3.38} & \textbf{\textit{0.9599 $\pm$ 0.0075}} & \textbf{\textit{0.0857 $\pm$ 0.0010}} \\
\midrule
Grep & Additional Greedy & 167.0 & 0.3577 & 0.4021 \\
 & SelectQA & 9.5 $\pm$ 0.53 & 0.3610 & 0.3079 \\
 & DIV-GA & 70.0 & 0.5790 $\pm$ 0.0227 & 0.2980 $\pm$ 0.0135 \\
 & QAOA-TCS & \textbf{419.2 $\pm$ 7.50} & \textbf{0.9384 $\pm$ 0.0157} & \textbf{0.0700 $\pm$ 0.0131} \\
 & \textit{QAOA-TCS (Ideal)} & \textit{326.2 $\pm$ 3.08} & \textit{0.7907 $\pm$ 0.0055} & \textit{0.0758 $\pm$ 0.0013} \\
\midrule
Gzip & Additional Greedy & 8.0 & 0.3332 & 0.4210 \\
 & SelectQA & 35.3 $\pm$ 0.82 & 0.4360 & 0.2464 \\
 & DIV-GA & \textbf{96.9 $\pm$ 6.74} & 0.6071 $\pm$ 0.0255 & 0.1938 $\pm$ 0.0152 \\
 & QAOA-TCS & 85.3 $\pm$ 19.59 & \textbf{0.7186 $\pm$ 0.0813} & \textbf{0.0931 $\pm$ 0.0193} \\
 & \textit{QAOA-TCS (Ideal)} & \textit{89.5 $\pm$ 0.53} & \textit{0.4828 $\pm$ 0.0031} & \textit{0.2567 $\pm$ 0.0025} \\
\midrule
Sed & Additional Greedy & 83.0 & 0.4776 & 0.2323 \\
 & SelectQA & 84.0 & 0.3907 & 0.3033 \\
 & DIV-GA & 99.2 $\pm$ 9.54 & 0.5845 $\pm$ 0.0140 & 0.2485 $\pm$ 0.0103 \\
 & QAOA-TCS & 115.8 $\pm$ 4.64 & 0.7143 $\pm$ 0.0075 & 0.0609 $\pm$ 0.0014 \\
 & \textit{QAOA-TCS (Ideal)} & \textbf{\textit{120.6 $\pm$ 3.10}} & \textbf{\textit{0.7221 $\pm$ 0.0056}} & \textbf{\textit{0.0582 $\pm$ 0.0016}} \\
\bottomrule
\end{tabular}%
\end{table*}

\begin{table*}[ht]
\centering
\scriptsize
\setlength{\tabcolsep}{3pt}
\caption{Mean and standard deviation of non-dominated solutions, HV, and IGD for QAOA-TCS compared to state-of-the-art methods on the Fake Brisbane backend with sampling noise. The italic row reports the Ideal QAOA-TCS reference.}
\label{tab:fakebris_means}
\begin{tabular}{llccc}
\toprule
\textbf{Program} & \textbf{Algorithm} & \textbf{Non-dom. sol.} $\uparrow$ & \textbf{HV} $\uparrow$ & \textbf{IGD} $\downarrow$ \\
\midrule
Flex & Additional Greedy & 10.0 & 0.3996 & 0.3455 \\
 & SelectQA & 187.0 & 0.4622 & 0.3237 \\
 & DIV-GA & 136.9 $\pm$ 3.38 & 0.6668 $\pm$ 0.0103 & 0.2147 $\pm$ 0.0087 \\
 & QAOA-TCS & \textbf{278.7 $\pm$ 70.19} & 0.8198 $\pm$ 0.2104 & 0.1336 $\pm$ 0.0294 \\
 & \textit{QAOA-TCS (Ideal)} & \textit{229.1 $\pm$ 3.38} & \textbf{\textit{0.9599 $\pm$ 0.0075}} & \textbf{\textit{0.0857 $\pm$ 0.0010}} \\
\midrule
Grep & Additional Greedy & 167.0 & 0.3577 & 0.3991 \\
 & SelectQA & 9.5 $\pm$ 0.53 & 0.3610 & 0.3097 \\
 & DIV-GA & 70.0 & 0.5790 $\pm$ 0.0227 & 0.2939 $\pm$ 0.0136 \\
 & QAOA-TCS & \textbf{419.8 $\pm$ 11.14} & \textbf{0.9427 $\pm$ 0.0079} & \textbf{0.0589 $\pm$ 0.0029} \\
 & \textit{QAOA-TCS (Ideal)} & \textit{326.2 $\pm$ 3.08} & \textit{0.7907 $\pm$ 0.0055} & \textit{0.0758 $\pm$ 0.0013} \\
\midrule
Gzip & Additional Greedy & 3.0 & 0.3332 & 0.4225 \\
 & SelectQA & 35.3 $\pm$ 0.82 & 0.4360 & 0.2456 \\
 & DIV-GA & \textbf{104.0 $\pm$ 1.15} & 0.6071 $\pm$ 0.0255 & 0.1865 $\pm$ 0.0149 \\
 & QAOA-TCS & 83.2 $\pm$ 16.73 & \textbf{0.7090 $\pm$ 0.0685} & \textbf{0.0934 $\pm$ 0.0183} \\
 & \textit{QAOA-TCS (Ideal)} & \textit{89.5 $\pm$ 0.53} & \textit{0.4828 $\pm$ 0.0031} & \textit{0.2567 $\pm$ 0.0025} \\
\midrule
Sed & Additional Greedy & 83.0 & 0.4779 & 0.2332 \\
 & SelectQA & 79.0 & 0.3909 & 0.3046 \\
 & DIV-GA & 99.2 $\pm$ 9.54 & 0.5845 $\pm$ 0.0140 & 0.2469 $\pm$ 0.0103 \\
 & QAOA-TCS & 118.2 $\pm$ 3.91 & 0.7196 $\pm$ 0.0104 & 0.0610 $\pm$ 0.0025 \\
 & \textit{QAOA-TCS (Ideal)} & \textbf{\textit{120.6 $\pm$ 3.10}} & \textbf{\textit{0.7221 $\pm$ 0.0056}} & \textbf{\textit{0.0582 $\pm$ 0.0016}} \\
\bottomrule
\end{tabular}%
\end{table*}

\begin{table*}[ht]
\centering
\scriptsize
\setlength{\tabcolsep}{3pt}
\caption{Mean and standard deviation of non-dominated solutions, HV, and IGD for QAOA-TCS compared with state-of-the-art methods under 1\% depolarizing error with sampling noise. The italic row reports the Ideal QAOA-TCS reference.}
\label{tab:noise1_means}
\begin{tabular}{llccc}
\toprule
\textbf{Program} & \textbf{Algorithm} & \textbf{Non-dom. sol.} $\uparrow$ & \textbf{HV} $\uparrow$ & \textbf{IGD} $\downarrow$ \\
\midrule
Flex & Additional Greedy & 1.0 & 0.3996 & 0.3183 \\
 & SelectQA & 187.0 & 0.4622 & 0.3535 \\
 & DIV-GA & 134.0 $\pm$ 7.26 & 0.6668 $\pm$ 0.0103 & 0.2154 $\pm$ 0.0084 \\
 & QAOA-TCS & \textbf{254.8 $\pm$ 63.44} & 0.7412 $\pm$ 0.2061 & 0.1403 $\pm$ 0.0043 \\
 & \textit{QAOA-TCS (Ideal)} & \textit{229.1 $\pm$ 3.38} & \textbf{\textit{0.9599 $\pm$ 0.0075}} & \textbf{\textit{0.0857 $\pm$ 0.0010}} \\
\midrule
Grep & Additional Greedy & 167.0 & 0.3577 & 0.4002 \\
 & SelectQA & 9.5 $\pm$ 0.53 & 0.3610 & 0.3147 \\
 & DIV-GA & 69.9 $\pm$ 0.32 & 0.5790 $\pm$ 0.0227 & 0.2905 $\pm$ 0.0136 \\
 & QAOA-TCS & \textbf{424.0 $\pm$ 10.24} & \textbf{0.9471 $\pm$ 0.0069} & \textbf{0.0541 $\pm$ 0.0014} \\
 & \textit{QAOA-TCS (Ideal)} & \textit{326.2 $\pm$ 3.08} & \textit{0.7907 $\pm$ 0.0055} & \textit{0.0758 $\pm$ 0.0013} \\
\midrule
Gzip & Additional Greedy & 8.0 & 0.3325 & 0.4299 \\
 & SelectQA & 35.3 $\pm$ 0.82 & 0.4360 & 0.2411 \\
 & DIV-GA & \textbf{101.0 $\pm$ 4.45} & 0.6071 $\pm$ 0.0255 & 0.1824 $\pm$ 0.0140 \\
 & QAOA-TCS & 88.9 $\pm$ 14.64 & \textbf{0.7357 $\pm$ 0.0658} & \textbf{0.0821 $\pm$ 0.0210} \\
 & \textit{QAOA-TCS (Ideal)} & \textit{89.5 $\pm$ 0.53} & \textit{0.4828 $\pm$ 0.0031} & \textit{0.2567 $\pm$ 0.0025} \\
\midrule
Sed & Additional Greedy & 83.0 & 0.4759 & 0.2376 \\
 & SelectQA & 78.0 & 0.3880 & 0.3063 \\
 & DIV-GA & 99.2 $\pm$ 9.54 & 0.5852 $\pm$ 0.0140 & 0.2484 $\pm$ 0.0103 \\
 & QAOA-TCS & 117.1 $\pm$ 3.75 & 0.7210 $\pm$ 0.0090 & 0.0598 $\pm$ 0.0019 \\
 & \textit{QAOA-TCS (Ideal)} & \textbf{\textit{120.6 $\pm$ 3.10}} & \textbf{\textit{0.7221 $\pm$ 0.0056}} & \textbf{\textit{0.0582 $\pm$ 0.0016}} \\
\bottomrule
\end{tabular}%
\end{table*}

\begin{table*}[ht]
\centering
\scriptsize
\setlength{\tabcolsep}{3pt}
\caption{Mean and standard deviation of non-dominated solutions, HV, and IGD for QAOA-TCS compared to state-of-the-art methods on the 2\% depolarizing error with sampling noise. The italic row reports the Ideal QAOA-TCS reference.}
\label{tab:noise2_means}
\begin{tabular}{llccc}
\toprule
\textbf{Program} & \textbf{Algorithm} & \textbf{Non-dom. sol.} $\uparrow$ & \textbf{HV} $\uparrow$ & \textbf{IGD} $\downarrow$ \\
\midrule
Flex & Additional Greedy & 8.0 & 0.3995 & 0.3155 \\
 & SelectQA & 187.0 & 0.4622 & 0.3543 \\
 & DIV-GA & 137.4 $\pm$ 3.44 & 0.6668 $\pm$ 0.0103 & 0.2156 $\pm$ 0.0085 \\
 & QAOA-TCS & \textbf{253.1 $\pm$ 67.65} & 0.7339 $\pm$ 0.2104 & 0.1429 $\pm$ 0.0029 \\
 & \textit{QAOA-TCS (Ideal)} & \textit{229.1 $\pm$ 3.38} & \textbf{\textit{0.9599 $\pm$ 0.0075}} & \textbf{\textit{0.0857 $\pm$ 0.0010}} \\
\midrule
Grep & Additional Greedy & 167.0 & 0.3577 & 0.3985 \\
 & SelectQA & 9.5 $\pm$ 0.53 & 0.3610 & 0.3084 \\
 & DIV-GA & 70.0 & 0.5790 $\pm$ 0.0227 & 0.2947 $\pm$ 0.0135 \\
 & QAOA-TCS & \textbf{422.4 $\pm$ 10.09} & \textbf{0.9396 $\pm$ 0.0055} & \textbf{0.0582 $\pm$ 0.0048} \\
 & \textit{QAOA-TCS (Ideal)} & \textit{326.2 $\pm$ 3.08} & \textit{0.7907 $\pm$ 0.0055} & \textit{0.0758 $\pm$ 0.0013} \\
\midrule
Gzip & Additional Greedy & 8.0 & 0.3332 & 0.4132 \\
 & SelectQA & 36.3 $\pm$ 0.82 & 0.4360 & 0.2548 \\
 & DIV-GA & \textbf{102.3 $\pm$ 2.50} & 0.6071 $\pm$ 0.0255 & 0.1910 $\pm$ 0.0160 \\
 & QAOA-TCS & 75.8 $\pm$ 19.05 & \textbf{0.6848 $\pm$ 0.0748} & \textbf{0.0978 $\pm$ 0.0131} \\
 & \textit{QAOA-TCS (Ideal)} & \textit{89.5 $\pm$ 0.53} & \textit{0.4828 $\pm$ 0.0031} & \textit{0.2567 $\pm$ 0.0025} \\
\midrule
Sed & Additional Greedy & 83.0 & 0.4744 & 0.2289 \\
 & SelectQA & 87.0 & 0.3868 & 0.2992 \\
 & DIV-GA & 99.2 $\pm$ 9.54 & 0.5831 $\pm$ 0.0140 & 0.2560 $\pm$ 0.0103 \\
 & QAOA-TCS & 117.2 $\pm$ 4.76 & 0.7128 $\pm$ 0.0075 & 0.0614 $\pm$ 0.0022 \\
 & \textit{QAOA-TCS (Ideal)} & \textbf{\textit{120.6 $\pm$ 3.10}} & \textbf{\textit{0.7221 $\pm$ 0.0056}} &\textbf{ \textit{0.0582 $\pm$ 0.0016}} \\
\bottomrule
\end{tabular}%
\end{table*}

\begin{table*}[t]
\centering
\scriptsize
\setlength{\tabcolsep}{3pt}
\caption{Mean and standard deviation of non-dominated solutions, HV, and IGD for QAOA-TCS compared with state-of-the-art methods under 5\% depolarizing error with sampling noise. The italic row reports the Ideal QAOA-TCS reference.}
\label{tab:noise5_means}
\begin{tabular}{llccc}
\toprule
\textbf{Program} & \textbf{Algorithm} & \textbf{Non-dom. sol.} $\uparrow$ & \textbf{HV} $\uparrow$ & \textbf{IGD} $\downarrow$ \\
\midrule
Flex & Additional Greedy & 1.0 & 0.3995 & 0.3580 \\
 & SelectQA & 187.0 & 0.4622 & 0.3145 \\
 & DIV-GA & 136.0 $\pm$ 4.52 & 0.6668 $\pm$ 0.0103 & 0.2158 $\pm$ 0.0087 \\
 & QAOA-TCS & \textbf{278.0 $\pm$ 64.96} & 0.8573 $\pm$ 0.1927 & 0.1242 $\pm$ 0.0366 \\
 & \textit{QAOA-TCS (Ideal)} & \textit{229.1 $\pm$ 3.38} & \textbf{\textit{0.9599 $\pm$ 0.0075}} & \textbf{\textit{0.0857 $\pm$ 0.0010}} \\
\midrule
Grep & Additional Greedy & 167.0 & 0.3577 & 0.4027 \\
 & SelectQA & 9.5 $\pm$ 0.53 & 0.3610 & 0.3216 \\
 & DIV-GA & 69.1 $\pm$ 1.29 & 0.5790 $\pm$ 0.0227 & 0.2864 $\pm$ 0.0137 \\
 & QAOA-TCS & \textbf{428.0 $\pm$ 8.31} & \textbf{0.9476 $\pm$ 0.0068} & \textbf{0.0530 $\pm$ 0.0045} \\
 & \textit{QAOA-TCS (Ideal)} & \textit{326.2 $\pm$ 3.08} & \textit{0.7907 $\pm$ 0.0055} & \textit{0.0758 $\pm$ 0.0013} \\
\midrule
Gzip & Additional Greedy & 8.0 & 0.3325 & 0.4200 \\
 & SelectQA & 35.3 $\pm$ 0.82 & 0.4360 & 0.2499 \\
 & DIV-GA & \textbf{104.1 $\pm$ 1.20} & 0.6071 $\pm$ 0.0255 & 0.1838 $\pm$ 0.0150 \\
 & QAOA-TCS & 81.4 $\pm$ 17.48 & \textbf{0.7067 $\pm$ 0.0605} & \textbf{0.0967 $\pm$ 0.0154} \\
 & \textit{QAOA-TCS (Ideal)} & \textit{89.5 $\pm$ 0.53} & \textit{0.4828 $\pm$ 0.0031} & \textit{0.2567 $\pm$ 0.0025} \\
\midrule
Sed & Additional Greedy & 83.0 & 0.4743 & 0.2376 \\
 & SelectQA & 78.0 & 0.3868 & 0.3046 \\
 & DIV-GA & 99.2 $\pm$ 9.54 & 0.5830 $\pm$ 0.0140 & 0.2467 $\pm$ 0.0103 \\
 & QAOA-TCS & 118.7 $\pm$ 3.80 & 0.7211 $\pm$ 0.0117 & 0.0606 $\pm$ 0.0028 \\
 & \textit{QAOA-TCS (Ideal)} & \textbf{\textit{120.6 $\pm$ 3.10}} & \textbf{\textit{0.7221 $\pm$ 0.0056}} & \textbf{\textit{0.0582 $\pm$ 0.0016}} \\
\bottomrule
\end{tabular}%
\end{table*}

Analyzing the results obtained on the sampling-noise-only simulator, as shown in Table~\ref{tab:samplenoise_means}, we observed that noise did not substantially alter the overall behavior of \textsc{QAOA-TCS}, since it can still perform better than the baseline approaches in \textit{Flex}, \textit{Grep}, and \textit{Sed}. The statistical analysis confirmed this trend with large effect sizes in most comparisons. In \textit{Gzip}, which is the only program where \textsc{DIV-GA} performs better than \textsc{QAOA-TCS}, the result is not statistically significant in terms of the number of non-dominated solutions; furthermore, \textsc{QAOA-TCS} performs better than \textsc{DIV-GA} in HV and IGD, and such findings are statistically significant with a large effect size. Noise not only does not alter the performance of \textsc{QAOA-TCS}, but the approach also performs better in \textit{Flex} and \textit{Grep}, while in \textit{Gzip} it achieves better HV and IGD results.

We observed a similar consistent pattern when analyzing the results from executing \textsc{QAOA-TCS} on the other noise simulators. In all such executions, \textsc{QAOA-TCS} still performs better than the baselines in \textit{Flex}, \textit{Grep}, and \textit{Sed} across all metrics, with statistical significance in most cases.

The strong performance of \textsc{QAOA-TCS} under noisy configurations can be explained by the fact that QAOA runs only on small clusters, each representing a subproblem of the original optimization problem. Even if the ideal simulator identifies the best solution for each subproblem, combining these locally optimal solutions does not necessarily yield the best global solution when merged to construct the final solution for the whole problem. Therefore, noise may introduce variations in the selected local solutions that, after the merging phase, can still lead to competitive or even better global trade-offs. Finally, since each quantum subproblem is small, the resulting circuits are relatively shallow, which reduces both the probability that noise affects the computation and the magnitude of its impact on the final solution.

\geribox{Takeaway \#3.}{\textsc{QAOA-TCS} is robust to noisy quantum environments and performs better than baseline approaches in three of four programs, with statistical analysis confirming the trend in most cases. Although on \textit{Gzip} \textsc{DIV-GA} appears to be better than \textsc{QAOA-TCS} in terms of the number of non-dominated solutions, these results are not statistically significant.}

\subsubsection{RQ$_{4.1}$ How efficient is \textsc{QAOA-TCS} compared with 
state-of-the-art approaches in its multi-objective formulation?}

\begin{table}[ht]
\centering
\scriptsize
\caption{Execution time in seconds for each algorithm across the analyzed programs.}
\label{tab:execution_time_by_program}
\begin{tabular}{llrr}
\toprule
\textbf{Program} & \textbf{Algorithm} & \textbf{Mean (s)} & \textbf{Std. Dev. (s)} \\
\midrule

\multirow{9}{*}{\textit{Flex}}
& \textbf{SelectQA}                    & \textbf{3.00}   & $<0.01$ \\
& DIV-GA                               & 186.84          & 18.00   \\
& Add-Greedy                           & 8.09            & $<0.01$ \\
& QAOA-TCS (Ideal simulator)           & 20.46           & 0.34    \\
& QAOA-TCS (Sampling Noise)            & 91.94           & 16.97   \\
& QAOA-TCS (Fake Brisbane)             & 2133.57         & 21.88   \\
& QAOA-TCS (Depol.\ 1\%)              & 46.54           & 1.57    \\
& QAOA-TCS (Depol.\ 2\%)              & 49.43           & 2.05    \\
& QAOA-TCS (Depol.\ 5\%)              & 48.02           & 1.58    \\
\midrule

\multirow{9}{*}{\textit{Grep}}
& \textbf{SelectQA}                    & \textbf{2.99}   & $<0.01$ \\
& DIV-GA                               & 69.91           & 9.39    \\
& Add-Greedy                           & 9.58            & $<0.01$ \\
& QAOA-TCS (Ideal simulator)           & 81.59           & 1.66    \\
& QAOA-TCS (Sampling Noise)            & 61.26           & 0.88    \\
& QAOA-TCS (Fake Brisbane)             & 2660.39         & 26.61   \\
& QAOA-TCS (Depol.\ 1\%)              & 59.35           & 1.48    \\
& QAOA-TCS (Depol.\ 2\%)              & 62.40           & 0.32    \\
& QAOA-TCS (Depol.\ 5\%)              & 61.29           & 0.66    \\
\midrule

\multirow{9}{*}{\textit{Gzip}}
& \textbf{Add-Greedy}                  & \textbf{0.27}   & $<0.01$ \\
& SelectQA                             & 2.99            & $<0.01$ \\
& DIV-GA                               & 11.87           & 1.56    \\
& QAOA-TCS (Ideal simulator)           & 16.46           & 0.78    \\
& QAOA-TCS (Depol.\ 1\%)              & 22.37           & 0.41    \\
& QAOA-TCS (Depol.\ 5\%)              & 23.49           & 0.38    \\
& QAOA-TCS (Sampling Noise)            & 24.11           & 0.80    \\
& QAOA-TCS (Depol.\ 2\%)              & 24.31           & 0.52    \\
& QAOA-TCS (Fake Brisbane)             & 1159.42         & 20.18   \\
\midrule

\multirow{9}{*}{\textit{Sed}}
& \textbf{Add-Greedy}                  & \textbf{2.09}   & $<0.01$ \\
& SelectQA                             & 2.99            & $<0.01$ \\
& DIV-GA                               & 57.97           & 2.57    \\
& QAOA-TCS (Depol.\ 1\%)              & 30.09           & 0.62    \\
& QAOA-TCS (Depol.\ 2\%)              & 32.20           & 0.56    \\
& QAOA-TCS (Sampling Noise)            & 32.24           & 0.54    \\
& QAOA-TCS (Depol.\ 5\%)              & 31.45           & 0.59    \\
& QAOA-TCS (Ideal simulator)           & 34.76           & 2.38    \\
& QAOA-TCS (Fake Brisbane)             & 1488.14         & 20.13   \\

\bottomrule
\end{tabular}
\end{table}

As shown in Table~\ref{tab:execution_time_by_program}, \textsc{SelectQA} and \textsc{Additional Greedy} are the fastest baselines, and statistical tests confirm this trend in almost all cases. For \textit{Flex} and \textit{Grep}, \textsc{SelectQA} is significantly faster than \textsc{DIV-GA} and almost all \textsc{QAOA-TCS} configurations, except for the depolarizing-noise ones. Looking at \textit{Gzip}, \textsc{Additional Greedy} is the most efficient approach, although no statistically significant difference was found in the comparison with \textsc{SelectQA}. Finally, on \textit{Sed}, \textsc{Additional Greedy} performed better than \textsc{DIV-GA} and \textsc{QAOA-TCS}, but without statistically relevant differences compared to \textsc{SelectQA}. 
However, these results should be interpreted cautiously since, unlike \textsc{QAOA-TCS}, \textsc{DIV-GA}, and \textsc{Additional Greedy}, which were executed on the same local machine, \textsc{SelectQA} relied on D-Wave's infrastructure. Therefore, external factors such as QPU availability, resource queues, and classical-quantum communication overhead may have affected its execution time.

\geribox{Takeaway \#4.}{\textsc{QAOA-TCS} appears to be less efficient than the \textsc{Additional Greedy} and \textsc{SelectQA} baselines, whose performance remained under ten seconds in each case.}

\subsection{Experimental Results of Single-Objective Algorithms}
This section presents the results of the comparison among all considered single-objective algorithms.

\subsubsection{RQ$_{1.2}$. How does the circuit depth of QAOA influence the effectiveness of QAOA-TCS in its single-objective formulation?}

To determine the best QAOA circuit depth, we ran \textsc{QAOA-TCS} ten times for each dataset at each depth configuration on the Ideal simulator and statistically compared the resulting $f$ values.

Regarding the \textit{GSDTSR} dataset, the results indicated that the optimal circuit depth is $p=1$. Regarding the other datasets, no statistically significant differences in the $f$ values have been observed, indicating that depth is not a determining factor in the quality of \textsc{QAOA-TCS} results. For computational cost reasons, we chose $p=1$ as the optimal setting also in this case.

\geribox{Takeaway \#5.}{Our analysis did not find statistically significant differences in result quality when varying the QAOA depth, except for \textit{GSDTSR}, where the best configuration was $p=1$. This result is similar to that obtained in the multi-objective formulation.}

\begin{table}[ht]
\centering
\caption{Execution cost and effectiveness of the final test suites produced by ideal \textsc{QAOA-TCS} and ideal \textsc{IGDec-QAOA} across \textit{GSDTSR}, \textit{PaintControl}, and \textit{IOF/ROL}.}
\label{tab:all_datasets_metrics_nobackend}
\scriptsize
\begin{tabular}{llrrrr}
\toprule
\textbf{Dataset} & \textbf{Algorithm} &
\multicolumn{2}{c}{\textbf{Execution Cost}} &
\multicolumn{2}{c}{\textbf{Effectiveness}} \\
\cmidrule(lr){3-4}\cmidrule(lr){5-6}
 & & \textbf{Mean} & \textbf{Std. Dev.} & \textbf{Mean} & \textbf{Std. Dev.} \\
\midrule

\textit{GSDTSR} &
\shortstack[l]{QAOA-TCS\\IGDec-QAOA} &
\shortstack[r]{\textbf{4,760,360.115}\\7,011,764.235} &
\shortstack[r]{0.000\\314,443.248} &
\shortstack[r]{7.571\\\textbf{7.804}} &
\shortstack[r]{0.000\\0.112} \\[4pt]

\textit{PaintControl} &
\shortstack[l]{QAOA-TCS\\IGDec-QAOA} &
\shortstack[r]{1,165,500.000\\\textbf{474,543.400}} &
\shortstack[r]{0.000\\130,380.215} &
\shortstack[r]{\textbf{14.455}\\8.382} &
\shortstack[r]{0.000\\0.477} \\[4pt]

\textit{IOF/ROL} &
\shortstack[l]{QAOA-TCS\\IGDec-QAOA} &
\shortstack[r]{64,190,011.000\\\textbf{10,801,269.300}} &
\shortstack[r]{0.000\\10,957,484.553} &
\shortstack[r]{\textbf{435.101}\\91.852} &
\shortstack[r]{0.000\\33.548} \\
\bottomrule
\end{tabular}
\end{table}

\begin{table}[ht]
\centering
\caption{Execution cost and input diversity for final suites of ideal \textsc{QAOA-TCS} and ideal \textsc{IGDec-QAOA} on \textit{ELEVATOR$_{o2}$}.}
\label{tab:elevator_o2_metrics_nobackend}
\scriptsize
\begin{tabular}{lrrrr}
\toprule
\textbf{Algorithm} &
\multicolumn{2}{c}{\textbf{Execution Cost}} &
\multicolumn{2}{c}{\textbf{Input Diversity}} \\
\cmidrule(lr){2-3}\cmidrule(lr){4-5}
 & \textbf{Mean} & \textbf{Std. Dev.} & \textbf{Mean} & \textbf{Std. Dev.} \\
\midrule
QAOA-TCS   & \textbf{133,136.000} & 0.000     & 722.161 & 0.000 \\
IGDec-QAOA & 171,895.600 & 19,856.754 & \textbf{751.797} & 26.447 \\
\bottomrule
\end{tabular}
\end{table}

\begin{table}[ht]
\centering
\caption{Execution cost, passenger count, and travel distance for ideal \textsc{QAOA-TCS} and ideal \textsc{IGDec-QAOA} on \textit{ELEVATOR$_{o3}$}.}
\label{tab:elevator_o3_metrics_nobackend}
\scriptsize
\begin{tabular}{lrrrr}
\toprule
\textbf{Algorithm} &
\multicolumn{2}{c}{\textbf{Execution Cost}} &
\multicolumn{2}{c}{\textbf{Passenger Count}} \\
\cmidrule(lr){2-3}\cmidrule(lr){4-5}
 & \textbf{Mean} & \textbf{Std. Dev.} & \textbf{Mean} & \textbf{Std. Dev.} \\
\midrule
QAOA-TCS   & \textbf{151,174.000} & 0.000    & 15,583.000 & 0.000 \\
IGDec-QAOA & 228,366.400 & 6,974.035 & \textbf{18,085.500} & 332.167 \\
\midrule
\multicolumn{1}{l}{} & \multicolumn{2}{c}{\textbf{Travel Distance}} & \multicolumn{2}{l}{} \\
\cmidrule(lr){2-3}
\multicolumn{1}{l}{} & \textbf{Mean} & \textbf{Std. Dev.} & \multicolumn{2}{l}{} \\
QAOA-TCS   & 50,652.000 & 0.000 & \multicolumn{2}{l}{} \\
IGDec-QAOA & \textbf{57,461.800} & 1,073.966 & \multicolumn{2}{l}{} \\
\bottomrule
\end{tabular}
\end{table}

\subsubsection{RQ$_{2.2}$.What is the effectiveness of \textsc{QAOA-TCS} compared with state-of-the-art approaches in its single-objective formulation when executed on an ideal simulator?}

As shown in Tables \ref{tab:all_datasets_metrics_nobackend} to \ref{tab:elevator_o3_metrics_nobackend}, when we evaluated \textsc{QAOA-TCS} against other single-objective approaches, neither approach consistently outperformed the other.
On \textit{GSDTSR}, \textsc{QAOA-TCS} found solutions with statistically significantly lower execution costs but lower effectiveness, although the latter was not statistically significant. The same behavior is also observed on \textit{IOF/ROL}, whereas on \textit{PaintControl} the opposite holds, as confirmed by statistical tests in both cases.
On \textit{ELEVATOR$_{o2}$}, \textsc{QAOA-TCS} finds solutions with lower execution costs, and statistical tests show significant advantages. Nevertheless, \textsc{IGDec-QAOA} obtained significantly better results in terms of effectiveness.
Finally, on \textit{ELEVATOR$_{o3}$}, \textsc{QAOA-TCS} finds better solutions in terms of execution costs, while \textsc{IGDec-QAOA} finds more effective subsuites. Such differences are also confirmed by statistical tests.

One advantage of both \textsc{QAOA-TCS} and \textsc{IGDec-QAOA} is that they let developers prioritize certain objectives over others by increasing the weight of a given objective. This approach is useful when software engineers must adapt the testing strategy to specific budgets or needs. For example, in contexts where safety is a primary requirement, testers may be more focused on fault detection and code coverage, rather than the total execution cost of the suite; on the contrary, in contexts where the time to market must be reduced, testers may want to focus on execution time, even at the cost of overlooking some faults.

\geribox{Takeaway \#6.}{The comparisons between \textsc{QAOA-TCS} and \textsc{IGDec-QAOA} executed on the ideal simulator did not find any approach outperforming the other. Specifically, except for \textit{PaintControl} and \textit{IOF/ROL}, \textsc{QAOA-TCS} finds subsuites with lower execution costs, whereas \textsc{IGDec-QAOA} finds solutions with higher effectiveness.}

\subsubsection{RQ$_{3.2}$. How robust is QAOA-TCS to noisy environments in its single-objective formulation?}

\begin{table}[ht]
\centering
\caption{Execution cost and effectiveness for QAOA-TCS under ideal and noisy backends, with IGDec-QAOA and baselines across \textit{GSDTSR}, \textit{PaintControl}, and \textit{IOF/ROL}.}
\label{tab:all_datasets_effweights_metrics}
\resizebox{\linewidth}{!}{
\begin{tabular}{lllrrrr}
\toprule
\textbf{Dataset} & \textbf{Algorithm} & \textbf{Backend} &
\multicolumn{2}{c}{\textbf{Execution Cost}} &
\multicolumn{2}{c}{\textbf{Effectiveness}} \\
\cmidrule(lr){4-5}\cmidrule(lr){6-7}
 & & & \textbf{Mean} & \textbf{Std. Dev.} & \textbf{Mean} & \textbf{Std. Dev.} \\
\midrule

\multirow{10}{*}{\textit{GSDTSR}} &
\multirow{6}{*}{QAOA-TCS}
  & Ideal simulator & 4,760,360.115 & 0.000       & 7.571 & 0.000 \\
& & Sampling Noise   & 8,718,030.344 & 854,123.279 & 7.556 & 0.076 \\
& & Fake Brisbane   & 9,415,096.470 & 1,228,184.813 & 7.570 & 0.087 \\
& & Depol. 1\%      & 8,581,464.059 & 898,137.807 & 7.510 & 0.104 \\
& & Depol. 2\%      & 9,958,825.121 & 1,084,539.609 & 7.567 & 0.117 \\
& & Depol. 5\%      & 9,917,680.314 & 755,688.405 & 7.516 & 0.150 \\
\cmidrule(lr){2-7}
& \multirow{2}{*}{IGDec-QAOA}
  & Ideal simulator & 7,011,764.235 & 314,443.248 & \textbf{7.804} & 0.112 \\
& & Noise           & 13,492,835.897 & 10,754,535.375 & 7.546 & 0.352 \\
\cmidrule(lr){2-7}
& SelectQA   & --    & \textbf{37,450.930} & 0.000       & 3.607 & 0.000 \\
& BootQA     & --    & 177,112.778   & 105,666.698  & 2.388 & 0.883 \\
\midrule

\multirow{10}{*}{\textit{PaintControl}} &
\multirow{6}{*}{QAOA-TCS}
  & Ideal simulator & 1,165,500.000 & 0.000      & 14.455 & 0.000 \\
& & Sampling Noise   & 1,280,915.500 & 133,928.657 & 14.254 & 0.768 \\
& & Fake Brisbane   & 1,324,022.600 & 117,556.078 & \textbf{14.535} & 0.694 \\
& & Depol. 1\%      & 1,249,780.500 & 90,883.107  & 14.089 & 0.548 \\
& & Depol. 2\%      & 1,266,441.900 & 132,615.689 & 14.095 & 0.767 \\
& & Depol. 5\%      & 1,333,351.200 & 147,559.808 & 14.334 & 0.904 \\
\cmidrule(lr){2-7}
& \multirow{2}{*}{IGDec-QAOA}
  & Ideal simulator & 474,543.400 & 130,380.215 & 8.382 & 0.477 \\
& & Noise           & 665,046.200 & 271,370.831 & 8.249 & 0.898 \\
\cmidrule(lr){2-7}
& SelectQA   & --    & \textbf{159,150.000} & 0.000      & 5.314 & 0.000 \\
& BootQA     & --    & 651,774.600 & 242,047.481 & 4.681 & 0.912 \\
\midrule

\multirow{8}{*}{\textit{IOF/ROL}} &
\multirow{6}{*}{QAOA-TCS}
  & Ideal simulator & 64,190,011.000 & 0.000 & \textbf{435.101} & 0.000 \\
& & Sampling Noise   & 67,106,136.200 & 1,080,994.454 & 431.113 & 3.199 \\
& & Fake Brisbane   & 66,306,510.700 & 1,146,640.452 & 427.149 & 3.836 \\
& & Depol. 1\%      & 66,468,837.300 & 677,719.930  & 428.841 & 1.977 \\
& & Depol. 2\%      & 67,198,194.600 & 673,543.891  & 429.711 & 2.361 \\
& & Depol. 5\%      & 67,147,794.900 & 1,293,047.297 & 429.386 & 4.357 \\
\cmidrule(lr){2-7}
& \multirow{2}{*}{IGDec-QAOA}
  & Ideal simulator & \textbf{10,801,269.300} & 10,957,484.553 & 91.852 & 33.548 \\
& & Noise           & 25,192,111.600 & 20,251,046.346 & 126.114 & 50.307 \\
\bottomrule
\end{tabular}}
\end{table}

\begin{table}[ht]
\centering
\caption{Execution cost and input diversity for QAOA-TCS under ideal and noisy backends, with IGDec-QAOA on \textit{Elevator}.}
\label{tab:elevator_effmetrics}
\scriptsize
\begin{tabular}{llrrrr}
\toprule
\textbf{Algorithm} & \textbf{Backend} & 
\multicolumn{2}{c}{\textbf{Execution Cost}} & 
\multicolumn{2}{c}{\textbf{Input Diversity}} \\
\cmidrule(lr){3-4} \cmidrule(lr){5-6}
 & & \textbf{Mean} & \textbf{Std. Dev.} & \textbf{Mean} & \textbf{Std. Dev.} \\
\midrule
\multirow{6}{*}{QAOA-TCS} 
  & Ideal simulator & \textbf{133,136.000} & 0.000    & 722.161 & 0.000 \\
  & Sampling Noise   & 134,086.900 & 567.045  & 721.297 & 0.827 \\
  & Fake Brisbane   & 134,097.700 & 680.828  & 721.502 & 1.484 \\
  & Depol. 1\%      & 134,541.200 & 617.643  & 722.590 & 0.875 \\
  & Depol. 2\%      & 134,450.000 & 759.870  & 722.288 & 1.565 \\
  & Depol. 5\%      & 134,798.400 & 1,043.149 & 722.275 & 1.676 \\
\midrule
\multirow{2}{*}{IGDec-QAOA}
  & Ideal simulator & 171,895.600 & 19,856.754 & \textbf{751.797} & 26.447 \\
  & Noise           & 185,283.700 & 17,829.767 & 724.662 & 63.954 \\
\bottomrule
\end{tabular}
\end{table}

\begin{table}[ht]
\centering
\caption{Execution cost, passenger count, and travel distance for QAOA-TCS under ideal and noisy backends, with IGDec-QAOA on \textit{Elevator-o3}.}
\label{tab:elevator2_effmetrics}
\scriptsize
\begin{tabular}{llrrrr}
\toprule
\textbf{Algorithm} & \textbf{Backend} &
\multicolumn{2}{c}{\textbf{Execution Cost}} &
\multicolumn{2}{c}{\textbf{Passenger Count}} \\
\cmidrule(lr){3-4}\cmidrule(lr){5-6}
 & & \textbf{Mean} & \textbf{Std. Dev.} & \textbf{Mean} & \textbf{Std. Dev.} \\
\midrule
\multirow{6}{*}{QAOA-TCS}
  & Ideal simulator & \textbf{151,174.000} & 0.000    & 15583.000 & 0.000 \\
  & Sampling Noise   & 154,704.600 & 1,743.549 & 15,355.300 & 89.187 \\
  & Fake Brisbane   & 156,858.200 & 1,579.536 & 15,402.500 & 86.557 \\
  & Depol. 1\%      & 155,394.400 & 2,851.423 & 15,354.900 & 139.885 \\
  & Depol. 2\%      & 156,584.700 & 1,581.175 & 15,383.200 & 83.629 \\
  & Depol. 5\%      & 158,309.000 & 1,377.001 & 15,446.400 & 76.969 \\
\midrule
\multirow{2}{*}{IGDec-QAOA}
  & Ideal simulator & 228,366.400 & 6,974.035 & \textbf{18,085.500} & 332.167 \\
  & Noise           & 226,640.400 & 19,202.746 & 17,298.200 & 1,579.883 \\
\midrule
 & & \multicolumn{2}{c}{\textbf{Travel Distance}} & & \\
\cmidrule(lr){3-4}
 & & \textbf{Mean} & \textbf{Std. Dev.} & & \\
\multirow{6}{*}{QAOA-TCS}
  & Ideal simulator & 50,652.000 & 0.000 & & \\
  & Sampling Noise   & 49,579.400 & 292.890 & & \\
  & Fake Brisbane   & 49,758.900 & 254.745 & & \\
  & Depol. 1\%      & 49,616.300 & 444.212 & & \\
  & Depol. 2\%      & 49,671.800 & 295.775 & & \\
  & Depol. 5\%      & 49,814.700 & 247.575 & & \\
\midrule
\multirow{2}{*}{IGDec-QAOA}
  & Ideal simulator & \textbf{57,461.800} & 1,073.966 & & \\
  & Noise           & 54,698.700 & 5,296.257 & & \\
\bottomrule
\end{tabular}
\end{table}

As shown in Tables \ref{tab:all_datasets_effweights_metrics} to \ref{tab:elevator2_effmetrics}, \textsc{QAOA-TCS} executed on noise simulators remains stable, with results similar to the ideal configuration and no statistically significant reductions in most cases. In \textit{PaintControl}, the Fake Brisbane configuration also outperformed the ideal one, achieving the best result in selecting the most effective subsuite.

On \textit{GSDTSR}, \textsc{SelectQA} achieved the most efficient final test suite, while \textsc{IGDec-QAOA} found the most effective solution, although neither result is statistically significant. On \textit{PaintControl}, \textsc{SelectQA} again found the most efficient solution, although the statistical analysis does not always support this; meanwhile, \textsc{QAOA-TCS} performed better on Fake Brisbane in terms of effectiveness. On \textit{IOF/ROL}, the results obtained in the ideal cases are confirmed also under noisy configurations, since noisy \textsc{IGDec-QAOA} found more efficient solutions, while noisy \textsc{QAOA-TCS} found more effective solutions; such results are always confirmed by the statistical tests.
The results are consistent with the ideal cases for the \textit{Elevator} datasets. In both $ELEVATOR_{o2}$ and $ELEVATOR_{o3}$, \textsc{QAOA-TCS} finds efficient solutions, while \textsc{IGDec-QAOA} finds effective solutions, with statistical analysis confirming both results.

The obtained results indicate that no clear winner emerges between the two QAOA-based approaches. Indeed, neither \textsc{QAOA-TCS} nor \textsc{IGDec-QAOA} consistently dominates the other across all metrics. For this reason, from an effectiveness perspective, neither approach is generally preferable.

\geribox{Takeaway \#7.}{The performance of single-objective \textsc{QAOA-TCS} executed on noisy environments is consistent with that obtained in the ideal simulator. However, no approach outperforms the others with \textsc{QAOA-TCS} generally finding more efficient subsuites, while \textsc{IGDec-QAOA} finds more effective subsuites.}

\subsubsection{RQ$_{4.2}$ How efficient is \textsc{QAOA-TCS} compared 
with state-of-the-art approaches in its single-objective formulation?}

\begin{table}[ht]
\centering
\caption{Execution time in seconds for each algorithm and backend across 
the analyzed datasets.}
\label{tab:execution_time_setting2}
\scriptsize
\setlength{\tabcolsep}{3pt}
\renewcommand{\arraystretch}{0.78}
\resizebox{\linewidth}{!}{
\begin{tabular}{lllrr}
\toprule
\textbf{Dataset} & \textbf{Algorithm} & \textbf{Backend} & 
\textbf{Mean (s)} & \textbf{Std. Dev. (s)} \\
\midrule

\multirow{10}{*}{\textit{GSDTSR}}
& \multirow{6}{*}{QAOA-TCS}
  & Ideal simulator   & 95.53   & 10.71   \\
& & Sampling Noise    & 9.31    & 0.91    \\
& & Fake Brisbane     & 2,035.92 & 20.24   \\
& & Depol.\ 1\%       & 17.71   & 0.39    \\
& & Depol.\ 2\%       & 18.93   & 4.59    \\
& & Depol.\ 5\%       & 17.71   & 0.27    \\
\cmidrule(lr){2-5}
& \multirow{2}{*}{IGDec-QAOA}
  & Ideal simulator   & 199.81  & 0.66    \\
& & Noise             & 1,557.85 & 99.35   \\
\cmidrule(lr){2-5}
& SelectQA  & --      & 3.00    & $<0.01$ \\
& \textbf{BootQA}    & --      & \textbf{0.58}    & 0.02    \\

\midrule
\multirow{10}{*}{\textit{PaintControl}}
& \multirow{6}{*}{QAOA-TCS}
  & Ideal simulator   & 27.72   & 0.21    \\
& & Sampling Noise    & 2.68    & 0.09    \\
& & Fake Brisbane     & 634.74  & 13.10   \\
& & Depol.\ 1\%       & 5.47    & 0.22    \\
& & Depol.\ 2\%       & 5.45    & 0.15    \\
& & Depol.\ 5\%       & 5.41    & 0.17    \\
\cmidrule(lr){2-5}
& \multirow{2}{*}{IGDec-QAOA}
  & \textbf{Ideal simulator}   & \textbf{0.03}    & $<0.01$ \\
& & Noise             & 0.25    & 0.02    \\
\cmidrule(lr){2-5}
& SelectQA  & --      & 2.99    & $<0.01$ \\
& BootQA    & --      & 0.91    & 0.04    \\

\midrule
\multirow{8}{*}{\textit{IOF/ROL}}
& \multirow{6}{*}{QAOA-TCS}
  & Ideal simulator   & 489.79  & 1.69    \\
& & Sampling Noise    & 54.67   & 1.98    \\
& & Fake Brisbane     & 11,378.77 & 163.47 \\
& & Depol.\ 1\%       & 108.28  & 2.45    \\
& & Depol.\ 2\%       & 108.34  & 3.01    \\
& & Depol.\ 5\%       & 109.74  & 5.42    \\
\cmidrule(lr){2-5}
& \multirow{2}{*}{IGDec-QAOA}
  & \textbf{Ideal simulator}   & \textbf{24.04}   & 0.11    \\
& & Noise             & 207.29  & 2.60    \\

\midrule
\multirow{8}{*}{\textit{$ELEVATOR_{o2}$}}
& \multirow{6}{*}{QAOA-TCS}
  & Ideal simulator   & 595.73  & 2.29    \\
& & Sampling Noise    & 84.02   & 2.18    \\
& & Fake Brisbane     & 22,844.96 & 116.05 \\
& & Depol.\ 1\%       & 153.97  & 4.28    \\
& & Depol.\ 2\%       & 151.39  & 4.16    \\
& & Depol.\ 5\%       & 150.65  & 3.65    \\
\cmidrule(lr){2-5}
& \multirow{2}{*}{IGDec-QAOA}
  & \textbf{Ideal simulator}   & \textbf{23.90}   & 0.66    \\
& & Noise             & 174.65  & 4.31    \\

\midrule
\multirow{8}{*}{\textit{$ELEVATOR_{o3}$}}
& \multirow{6}{*}{QAOA-TCS}
  & Ideal simulator   & 833.13  & 3.53    \\
& & Sampling Noise    & 83.58   & 2.35    \\
& & Fake Brisbane     & 20,891.92 & 80.24  \\
& & Depol.\ 1\%       & 148.04  & 4.58    \\
& & Depol.\ 2\%       & 148.17  & 3.42    \\
& & Depol.\ 5\%       & 149.12  & 4.57    \\
\cmidrule(lr){2-5}
& \multirow{2}{*}{IGDec-QAOA}
  & \textbf{Ideal simulator}   & \textbf{23.75}   & 0.13    \\
& & Noise             & 174.01  & 4.92    \\
\bottomrule
\end{tabular}}
\end{table}

Table~\ref{tab:execution_time_setting2} reveals that \textsc{BootQA} and the ideal \textsc{IGDec-QAOA} are the most efficient approaches, consistently achieving the lowest execution times across all datasets. On \textit{GSDTSR}, \textsc{BootQA} is statistically significantly faster than all \textsc{QAOA-TCS} approaches, while \textsc{IGDec-QAOA} is considerably slower than the depolarizing \textsc{QAOA-TCS} configurations. On \textit{PaintControl}, \textit{IOF/ROL}, \textit{$ELEVATOR_{o2}$}, and \textit{$ELEVATOR_{o3}$} ideal \textsc{IGDec-QAOA} is significantly faster than all \textsc{QAOA-TCS} configurations with large effect sizes, but, when executed on noisy environments, \textsc{IGDec-QAOA} on \textit{GSDTSR}, \textit{IOF/ROL}, \textit{$ELEVATOR_{o2}$}, and \textit{$ELEVATOR_{o3}$} has always worse results in terms of execution times than \textsc{QAOA-TCS}, except for the Fake Brisbane configuration; such observation is confirmed by the statistical tests with large effect sizes.

These results are interesting from a practitioner's perspective because, although they do not represent large-scale systems, the adopted datasets come from small- and medium-scale industrial systems. Indeed, for the small-scale \textit{PaintControl} dataset, all approaches and most \textsc{QAOA-TCS} configurations could select test cases in less than 6 seconds. Regarding \textit{IOF/ROL, $ELEVATOR_{o2}$}, and \textit{$ELEVATOR_{o3}$}, although ideal \textsc{IGDec-QAOA} was executed in less than 25 seconds, in a noisy environment, some configurations require more than two minutes to complete the task. Such considerations suggest that the execution time is strictly problem-dependent, since in the case of the largest dataset, i.e., \textsc{GSDTSR}, the execution times of most \textsc{QAOA-TCS} configurations solved the problem in less than 20 seconds, while \textsc{BootQA} was executed in less than 1 second. This suggests that on larger datasets, QAOA-based approaches may suffer from scalability issues, since current limits on the number of supported qubits force them to rely on a hybrid classical-quantum architecture, thereby worsening performance.

Given these efficiency results, we observed that, although on ideal simulators \textsc{IGDec-QAOA} performs much better than the other approaches, the situation is reversed when the QAOA-based approaches are executed on a realistic noisy environment, where \textsc{QAOA-TCS} performs the TCS task more efficiently than \textsc{IGDec-QAOA} in most cases. From this perspective, practitioners would choose \textsc{QAOA-TCS} over \textsc{IGDec-QAOA} only for efficiency. However, these results are hindered by the different computational environments used to run the approaches.
As in the multi-objective settings, \textsc{BootQA} and \textsc{SelectQA} have been executed on \textit{D-Wave}'s infrastructure, while \textsc{QAOA-TCS} and \textsc{IGDec-QAOA} have been executed on the same traditional local machine; therefore, these results should be interpreted cautiously.

\geribox{Takeaway \#8.}{From the efficiency perspective, \textsc{IGDec-QAOA} executed on the ideal simulator is the best-performing approach in most cases, but when executed on noisy environments, \textsc{QAOA-TCS} performs better than \textsc{IGDec-QAOA} in most cases.}

\section{Threats to Validity}
\label{sec:5}
This section discusses the main threats to the validity of our study. 

\paragraph{Construct Validity.}
This threat entails the process we followed to collect data and implement the algorithms. We relied on state-of-the-art open-source tools (e.g., GNU and gcov). Furthermore, where possible, we reused the baselines' original implementations~\citep{selectqa,panichella2014improving,bootqa,igdec_qaoa}. Finally, we provide all scripts we implemented in our online appendix~\citep{appendix} to enhance the verifiability of our results.

\paragraph{Internal Validity.}  
The stochastic nature of QAOA and QA introduces randomness in the outcomes. To reduce variance, we executed each configuration 10 times, in line with~\citet{selectqa,bootqa}.
The decomposition size ($N=7$) used for \textsc{QAOA-TCS} and \textsc{IGDec-QAOA} and the QAOA circuit depth ($p=1$) chosen for \textsc{IGDec-QAOA} constitute another potential threat, as different decomposition granularities and circuit depths may alter effectiveness. We selected these values consistent with prior work on \textsc{IGDec-QAOA}~\citep{igdec_qaoa} to enable fair comparisons between \textsc{QAOA-TCS} and \textsc{IGDec-QAOA}.
The parameters used in the QUBO formalizations pose a potential threat; therefore, we employed the Optuna framework to determine such values for \textsc{QAOA-TCS} and \textsc{IGDec-QAOA}, while all other approaches were configured using the parameter settings adopted in their original works.
Although execution time is a straightforward measure of efficiency, a further threat to internal validity is the different environments used to run the experiments. Indeed, while all state-of-the-art approaches, \textsc{QAOA-TCS} and \textsc{IGDec-QAOA}, have always been executed on the same machine, \textsc{SelectQA} and \textsc{BootQA} leveraged \textit{D-Wave}'s infrastructure. For this reason, the runtime of both QA approaches could have been influenced by external factors, such as QPU availability, resource queues, and classical-quantum communication overhead, limiting the comparability of the execution times.

\paragraph{External Validity.}
Our findings are based on executing \textsc{QAOA-TCS} on widely used datasets in the literature, particularly those used by the algorithms we compare with. However, a key concern is the generalizability of the results beyond the systems considered in this work, as their limited scale may not represent certain industrial scenarios.
Furthermore, although we have evaluated the robustness of \textsc{QAOA-TCS} on noisy simulators, these environments cannot fully replicate the conditions encountered on real quantum devices. In particular, phenomena such as calibration drift~\citep{yeter2023measuring}, correlated errors~\citep{harper2023}, and limited qubit connectivity~\citep{cheng2024}, which affect quantum algorithm performance, are hard to capture with noise models. 
Although simulators directly inspired by real quantum devices, such as Fake Brisbane, can account for many hardware characteristics, they cannot necessarily capture all these effects under actual operating conditions. Therefore, because the robustness observed under the considered noise models does not guarantee equivalent results on real quantum devices, future experiments should be performed on real quantum hardware.


\paragraph{Conclusion Validity.}  
A first threat concerns the choice of evaluation metrics; therefore, we have relied on metrics used in other well-established state-of-the-art work~\citep{selectqa,panichella2014improving,bootqa,igdec_qaoa}.
We applied non-parametric statistical tests for non-normal distributions and parametric tests for normal distributions. Finally, we used a well-established effect-size measure to quantify the magnitude of the statistically significant differences between algorithms. No conclusion was finalized without statistically grounded evidence.

\section{Conclusions and Future Work}
\label{sec:6}

In this work, we have investigated in depth the effectiveness of \textsc{QAOA-TCS}, our novel clustering-based QAOA approach for regression test case selection. In particular, we aimed to demonstrate its robustness to quantum noise by running additional experiments with noisy-environment simulators and to establish its advantages over both classical and state-of-the-art quantum approaches in effectiveness and efficiency.
Since state-of-the-art approaches are divided into multi- and single-objective approaches, and to allow a fair comparison between them and \textsc{QAOA-TCS}, we have developed two versions of our approach, one multi-objective, which builds a Pareto frontier from a single solution, the other single-objective, yielding a single sub-suite after each execution.

The results showed that multi-objective \textsc{QAOA-TCS} outperforms all the multi-objective state-of-the-art approaches in most cases and, more importantly, these results remain stable even when \textsc{QAOA-TCS} is subject to high levels of noise. In the single-objective formulation, no clear winner emerges among the compared approaches. Our previous QA approach, namely \textsc{SelectQA}, has been shown to be the most efficient for large problems in the multi-objective setting, while \textsc{QAOA-TCS} offers the best trade-off between effectiveness and efficiency in the single-objective setting when executed in realistic noisy environments. These findings justify further experiments using real quantum systems, which are left for future work. They would also allow us to fully evaluate the effectiveness of noise-mitigation strategies, such as zero-noise extrapolation, dynamical decoupling, and others, in the context of regression test case selection.

From the practitioners' perspective, the main reason for choosing \textsc{QAOA-TCS} over \textsc{IGDec-QAOA} and Quantum Annealing is its higher efficiency in realistic noisy environments, since all quantum-based approaches achieved comparable effectiveness. Nevertheless, in the multi-objective case, \textsc{SelectQA}, although being outperformed by \textsc{QAOA-TCS} in terms of effectiveness, was significantly better than \textsc{QAOA-TCS} in terms of efficiency, suggesting that the computational costs of each algorithm in solving the TCS problem are highly problem-dependent, making it challenging to establish a clear winner in terms of efficiency.

In addition to demonstrating the robustness of QAOA-TCS under different noise models, our results indicate that noise does not necessarily degrade performance monotonically. In several cases, noisy executions yielded Pareto-front indicators comparable to or even better than those obtained on the ideal simulator. While previous studies have suggested that stochasticity and specific noise structures can occasionally benefit variational optimization, the consistency of these patterns across multiple noise settings suggests additional mechanisms may be at play. In particular, the interaction between noise-induced diversification and the decomposition-recombination strategy employed by QAOA-TCS deserves further investigation. Future work will therefore focus on systematically studying how noise, decomposition strategies, and problem structure affect the quality of the resulting Pareto frontiers, as well as on developing more robust decomposed variational approaches for software engineering applications.

\section{Declarations}

\subsection{Ethical approval}
Ethical approval: not applicable.

\subsection{Informed consent}
Informed consent: not applicable.

\subsection{Data Availability Statement}
The authors confirm that the data supporting the findings of this study are available at \url{https://zenodo.org/records/20668761}.

\subsection{Conflict of interest}
The authors declare no conflicts of interest.

\subsection{Clinical Trial Number}
Clinical Trial Number: not applicable.

\subsection{Funding}
This work was supported by the European Union - Next Generation EU under Mission 4, Component 1 (CUP D53D23017570001 and E53D23016370001), and by the Italian Government (Ministry of University and Research) through the PRIN 2022 PNRR project P2022SELA7, ``RECHARGE: monitoRing, tEsting, and CHaracterization of performAnce Regressions'' --  Decreto Direttoriale n. 1205, 28 July 2023.

\bibliographystyle{spbasic}   
\bibliography{references}     

\end{document}